\documentstyle[11pt,dina4]{article}

\begin{document}
\def\gsim{\hbox{$\lower1pt\hbox{$>$}\above-1pt\raise1pt\hbox{$\sim$}$}}
\def\lsim{\hbox{$\lower1pt\hbox{$<$}\above-1pt\raise1pt\hbox{$\sim$}$}}
\title{\mbox{
Theory of magnetic short--range order for itinerant electron systems}}
\author{{\sc U. Trapper$^{(1)}$}, {\sc  D. Ihle$^{(1)}$}
and {\sc H.~Fehske$^{(2)}$} \\[0.5cm]
$^{(1)}$Institut f\"ur Theoretische Physik, Universit\"at Leipzig,
D--04109 Leipzig, Germany\\[0.3cm]
$^{(2)}$Physikalisches Institut, Universit\"at Bayreuth,
D--95440 Bayreuth, Germany\\[0.6cm]}
% \date{Bayreuth, August 26, 1996}
\maketitle
%% Definitionsfile fuer Formelabkuerzungen usw.
%%
%%
% Neue Befehle fuer Kap. und SEct. im Header
%------------------------------------------------------------------
\newcommand{\hchapter}{Kapitel \thechapter.~}
\newcommand{\hsection}{\thechapter.~}
%
%
% Formelnumerierung im array mit 2.2.a usw
%-------------------------------------------------------------------
%  \alpheqn :  einschalten
%  \reseteqn:  ausschalten
%-------------------------------------------------------------------
\newcounter{saveeqn}
\newcommand{\eqnnumber}{\setcounter{saveeqn}{\value{equation}}%
\addtocounter{saveeqn}{1}%
\setcounter{equation}{0}%
\renewcommand{\theequation}{\arabic{section}.\arabic{equation}}}
\newcommand{\appAeqn}{\setcounter{saveeqn}{\value{equation}}%
\addtocounter{saveeqn}{1}%
\setcounter{equation}{0}%
\renewcommand{\theequation}{A.\arabic{equation}}}
\newcommand{\appBeqn}{\setcounter{saveeqn}{\value{equation}}%
\addtocounter{saveeqn}{1}%
\setcounter{equation}{0}%
\renewcommand{\theequation}{B.\arabic{equation}}}
\newcommand{\appCeqn}{\setcounter{saveeqn}{\value{equation}}%
\addtocounter{saveeqn}{1}%
\setcounter{equation}{0}%
\renewcommand{\theequation}{C.\arabic{equation}}}
\newcommand{\reseteqn}{\setcounter{equation}{\value{saveeqn}}%
\renewcommand{\theequation}{\arabic{equation}}}
%
%
% Neue mathemaische Zeichen
%-------------------------------------------------------------------
%  \gapro :  groesser ungefaehr  >~
%  \lapro :  kleiner ungefaehr   <~
%-------------------------------------------------------------------
\newcommand{\gapro}
  {\raisebox{-0.25ex} {$\,\stackrel{\scriptscriptstyle>}%
    {\scriptscriptstyle\sim}\,$}}
\newcommand{\lapro}
   {\raisebox{-0.25ex} {$\,\stackrel{\scriptscriptstyle<}%
    {\scriptscriptstyle\sim}\,$}}
%
%
% Sonstiges
%-------------------------------------------------------------------
\def\DS{\displaystyle}
\def\SS{\scriptstyle}
%
% Caligraphische Buchstaben
%---------------------------
\def\cZ{{\cal Z}}
\def\cD{{\cal D}}
\def\cL{{\cal L}}
\def\cS{{\cal S}}
\def\cH{{\cal H}}
\def\cG{{\cal G}}
\def\cF{{\cal F}}
\def\cJ{{\cal J}}
\def\cE{{\cal E}}
\def\cP{{\cal P}}
%
%
%
%  Substanzen                               
\def\LSCO{$\rm La_{2-x}Sr_xCuO_4$}
\def\NCCO{$\rm Nd_{2-x}Ce_xCuO_4$}
\def\YBCO{$\rm YBa_2Cu_3O_{6+x}$}
%
%
%
%
%=====================================================================
%  Operatoren/Felder       (Platzindex i)
%=====================================================================
%  Fermi-Operatoren
%----------------------------------------
\def\cis{c_{i\sigma}^{ }}                   
\def\cisd{c_{i\sigma}^{\dagger}}
\def\cisc{c_{i\sigma}^{*}}
\def\cjs{c_{j\sigma}^{ }}
\def\cjsd{c_{j\sigma}^{\dagger}}
\def\cjsc{c_{j\sigma}^{*}}
\def\cisdcjs{\cisd\cjs}
\def\cisccjs{\cisc\cjs}
\def\cisdcis{\cisd\cis}
%----------------------------------------
%  Pseudo-Fermi-Operatoren
%----------------------------------------
\def\fis{f_{i\sigma}^{ }}                   
\def\fisd{f_{i\sigma}^{\dagger}}
\def\fisc{f_{i\sigma}^{*}}
\def\fjs{f_{j\sigma}^{ }}
\def\fjsd{f_{j\sigma}^{\dagger}}
\def\fjsc{f_{j\sigma}^{*}}
\def\fisdfjs{\fisd\fjs}
\def\fiscfjs{\fisc\fjs}
\def\fisdfis{\fisd\fis}
%----------------------------------------
%  Einfach besetzte Plaetze (Spin sigma) 
%----------------------------------------
\def\pis{p_{i\sigma}^{ }}
\def\pisd{p_{i\sigma}^{\dagger}}
\def\pisc{p_{i\sigma}^{*}}
\def\pjs{p_{j\sigma}^{ }}
\def\pjsd{p_{j\sigma}^{\dagger}}
\def\pjsc{p_{j\sigma}^{*}}
\def\pisdpjs{\pisd\pjs}
\def\piscpjs{\pisc\pjs}
\def\pisdpis{\pisd\pis}
\def\pispis{p_{i\sigma}^{2}}
%----------------------------------------
%  Einfach besetzte Plaetze (Spin -sigma) 
%----------------------------------------
\def\pims{p_{i-\sigma}^{ }}
\def\pimsd{p_{i-\sigma}^{\dagger}}
\def\pimsc{p_{i-\sigma}^{*}}
\def\pimsdpims{\pimsd\pims}
%----------------------------------------
%  leere Plaetze
%----------------------------------------
\def\ei{e_{i}^{ }}
\def\eid{e_{i}^{\dagger}}
\def\eic{e_{i}^{*}}
\def\ej{e_{j}^{ }}
\def\ejd{e_{j}^{\dagger}}
\def\ejc{e_{j}^{*}}
\def\eidei{\eid\ei}
\def\eicei{\eic\ei}
\def\eiei{e_{i}^{2}}
%----------------------------------------
%  doppelt besetzte Plaetze
%----------------------------------------
\def\di{d_{i}^{ }}
\def\did{d_{i}^{\dagger}}
\def\dic{d_{i}^{*}}
\def\dj{d_{j}^{ }}
\def\djd{d_{j}^{\dagger}}
\def\djc{d_{j}^{*}}
\def\diddi{\did\di}
\def\dicdi{\dic\di}
%-------------------------------
%  Renormierungsfaktor    
%-------------------------------
\def\zis{z_{i\sigma}^{ }}
\def\zisd{z_{i\sigma}^{\dagger}}
\def\zisc{z_{i\sigma}^{*}}
\def\zjs{z_{j\sigma}^{ }}
\def\zjsd{z_{j\sigma}^{\dagger}}
\def\zjsc{z_{j\sigma}^{*}}
\def\zisdzjs{\zisd\zjs}
\def\zisczjs{\zisc\zjs}
\def\zisczis{\zisc\zis}
\def\qis{q_{i\sigma}^{ }}
\def\qos{q_{\sigma}^{o}}
\def\qo{q_{ }^{o}}
%-------------------------------
%  Lagrange-Multiplikatoren   
%-------------------------------
\def\lbis{\lambda_{i\sigma}^{(2)}}
\def\lbi{\lambda_{i}^{(1)}}
\def\lbjs{\lambda_{i\sigma}^{(2)}}
\def\lbj{\lambda_{i}^{(1)}}
\def\lbos{\lambda_{\sigma}^{o}}
\def\lbo{\lambda_{ }^{o}}
%-------------------------------
%  Hase-transformierte Felder  
%-------------------------------
\def\ni{n_i^{ }}
\def\nui{\nu_i^{ }}
\def\mi{m_i^{ }}
\def\mb{\bar{m}}
\def\xii{\xi_i^{ }}
\def\xib{\bar{\xi}}
%
%
%  Ising-Spin
%--------------------------------------------------------------------------
\def\si{s_{i}^{ }}
\def\sj{s_{j}^{ }}
\def\sk{s_{k}^{ }}
\def\alpi{\alpha_i^{ }}
\def\alpj{\alpha_j^{ }}
\def\alpk{\alpha_k^{ }}
%
%
%
%
%=====================================================================
%  Operatoren/Felder       (Index s_i)
%=====================================================================
\def\dsi{d_{\si}^{ }}
\def\dsic{d_{\si}^{*}}
\def\nsi{n_{\si}^{ }}
\def\nusi{\nu_{\si}^{ }}
\def\mbsi{\mb_{\si}^{ }}
\def\xibsi{\xib_{\si}^{ }}
\def\dcdsi{\dsic\dsi}
%
%
%
%
%=====================================================================
%  Operatoren/Felder       (Index alpha)
%=====================================================================
\def\da{d_{\alpha}^{ }}
\def\dac{d_{\alpha}^{*}}
\def\na{n_{\alpha}^{ }}
\def\nua{\nu_{\alpha}^{ }}
\def\mba{\mb_{\alpha}^{ }}
\def\xiba{\xib_{\alpha}^{ }}
\def\dcda{\dac\da}
\def\lbas{\lambda_{\alpha\sigma}^{(2)}}
\def\qas{q_{\alpha\sigma}^{ }}
\def\lbet{\lambda_{\eta}^{(2)}}
\def\qet{q_{\eta}^{ }}
%
%
%
%
%  Quasiteilchenenergie, Banddispersion usw.
%--------------------------------------------------------------------------
\def\Eks{E_{\vec{k}\sigma}^o}
\def\kx{k_x^{ }}
\def\ky{k_y^{ }}
\def\ek{\varepsilon_{\vec{k}}}
\def\ts{t_{ }^{\prime}}
%
%
%
%  Fermi-Funktion
\def\fermi{f(\omega-\mu)}
\def\fermiEs{f(\qos\omega+\lbos-\sigma h-\mu)}
%
%
%
%
% Shiba, Stoerungsentwicklung
%-------------------------------------------------------------------------
\def\hG{\hat{G}_{ij\sigma}^{ }}
\def\ihG{\hat{G}_{ij\sigma}^{-1}}
\def\hGo{\hat{G}_{ij\sigma}^{o}}
\def\hGoii{\hat{G}_{ii\sigma}^{o}}
\def\hGoij{\hat{G}_{(ij)\sigma}^{o}}
\def\hGoji{\hat{G}_{(ji)\sigma}^{o}}
\def\ihGo{\hat{G}_{ij\sigma}^{o^{-1}}}
\def\ihGoii{\hat{G}_{ii\sigma}^{o^{-1}}}
\def\hGon{\hat{G}_{n,\sigma}^{o}}
\def\hGonn{\hat{G}_{n,\sigma}^{o^2}}
\def\hGso{\hat{G}_{0}^{o}}
\def\hGsnn{\hat{G}_{n}^{o^2}}
\def\Vis{V_{i\sigma}^{ }}
\def\Vjs{V_{j\sigma}^{ }}
\def\Tis{T_{i\sigma}^{ }}
\def\Tjs{T_{j\sigma}^{ }}
\def\Tas{T_{\alpha\sigma}^{ }}
\def\Tbs{T_{\alpha^{\prime}\sigma}^{ }}
\def\qaas{a_{\alpha\sigma}^{2}}
\def\aas{a_{\alpha\sigma}^{ }}
\def\aass{a_{\alpha'\sigma}^{ }}
\def\aasss{a_{\alpha''\sigma}^{ }}
\def\bas{b_{\alpha\sigma}^{ }}
\def\bass{b_{\alpha'\sigma}^{ }}
\def\basss{b_{\alpha''\sigma}^{ }}
\def\Vas{V_{\alpha\sigma}^{ }}
\def\Tas{T_{\alpha\sigma}^{ }}
\def\Tass{T_{\alpha'\sigma}^{ }}
\def\basl{b_{\alpha\sigma}^{l}}
\def\basls{b_{\alpha'\sigma}^{l'}}
\def\aet{a_{\eta}^{ }}
\def\aett{a_{\eta}^{2}}
\def\aettt{a_{\eta}^{3}}
\def\bet{b_{\eta}^{ }}
\def\bett{b_{\eta}^{2}}
\def\bettt{b_{\eta}^{3}}
\def\amet{a_{-\eta}^{ }}
\def\amett{a_{-\eta}^{2}}
\def\amettt{a_{-\eta}^{3}}
\def\bmet{b_{-\eta}^{ }}
\def\bmett{b_{-\eta}^{2}}
\def\bmettt{b_{-\eta}^{3}}
\def\Vet{V_{\eta}^{ }}
\def\Tet{T_{\eta}^{ }}
\def\Tets{T_{\eta'}^{ }}
\def\betl{b_{\eta}^{l}}
\def\batls{b_{\eta'}^{l'}}
\def\bPsi{\bar{\mit \Psi}}
\def\bh{\bar{h}}
\def\bJn{\bar{J}_{n}^{ }}
\def\bJone{\bar{J}_{1}^{ }}
\def\Kom{K_{\bh}^{ }}
\def\Knm{K_{\bJn}^{ }}
\def\Konem{K_{\bJone}^{ }}
%
%
%
% Gitter-Greens-Funktion
%-------------------------------------------------------------------------
\def\kxp{k_x^{\prime}}
\def\kyp{k_y^{\prime}}
\def\npr{n_{ }^{\prime}}
\def\mpr{m_{ }^{\prime}}
%
%
%

%% Definitionsfile fuer Abkuerzungen von Bildelementen  usw.
%%
%%
%--------------------------------------------
% Definition von Grafikelementen (klein) (vuer CVM)
%-------------------------------------------------------------------------
\unitlength0.3mm
%-------------------------------------------------------------------------
%       o-o-o
%       | | |
%       o-o-o   9-Punkt-Cluster
%       | | |
%       o-o-o
%--------------------------------------------
\newsavebox{\kneun}
\savebox{\kneun}{
  \begin{picture}(10,10)
    \put( 0, 0){\circle*{2.5}}
    \put( 0, 0){\line(1,0){5}}
    \put( 0, 0){\line(0,1){5}}
    \put( 0, 5){\circle*{2.5}}
    \put( 0, 5){\line(1,0){5}}
    \put( 0, 5){\line(0,1){5}}
    \put( 0,10){\circle*{2.5}}
    \put( 0,10){\line(1,0){5}}
    \put( 5, 0){\circle*{2.5}}
    \put( 5, 0){\line(1,0){5}}
    \put( 5, 0){\line(0,1){5}}
    \put( 5, 5){\circle*{2.5}}
    \put( 5, 5){\line(1,0){5}}
    \put( 5, 5){\line(0,1){5}}
    \put( 5,10){\circle*{2.5}}
    \put( 5,10){\line(1,0){5}}
    \put(10, 0 ){\circle*{2.5}}
    \put(10, 5){\circle*{2.5}}
    \put(10, 0){\line(0,1){5}}
    \put(10,10){\circle*{2.5}}
    \put(10, 5){\line(0,1){5}}
  \end{picture}
}
%--------------------------------------------
%       o-o-o
%       | | |    6-Punkt-Cluster
%       o-o-o   
%--------------------------------------------
\newsavebox{\ksechs}
\savebox{\ksechs}{
  \begin{picture}(10,10)
    \put( 0, 2.5){\circle*{2.5}}
    \put( 0, 2.5){\line(1,0){5}}
    \put( 0, 2.5){\line(0,1){5}}
    \put( 0, 7.5){\circle*{2.5}}
    \put( 0, 7.5){\line(1,0){5}}
    \put( 5, 2.5){\circle*{2.5}}
    \put( 5, 2.5){\line(1,0){5}}
    \put( 5, 2.5){\line(0,1){5}}
    \put( 5, 7.5){\circle*{2.5}}
    \put( 5, 7.5){\line(1,0){5}}
    \put(10, 2.5){\circle*{2.5}}
    \put(10, 7.5){\circle*{2.5}}
    \put(10, 2.5){\line(0,1){5}}
  \end{picture}
}
%--------------------------------------------
%       o-o
%       | |    4-Punkt-Cluster
%       o-o   
%--------------------------------------------
\newsavebox{\kvier}
\savebox{\kvier}{
  \begin{picture}(10,10)
    \put( 2.5, 2.5){\circle*{2.5}}
    \put( 2.5, 2.5){\line(1,0){5}}
    \put( 2.5, 2.5){\line(0,1){5}}
    \put( 2.5, 7.5){\circle*{2.5}}
    \put( 2.5, 7.5){\line(1,0){5}}
    \put( 7.5, 2.5){\circle*{2.5}}
    \put( 7.5, 2.5){\line(0,1){5}}
    \put( 7.5, 7.5){\circle*{2.5}}
  \end{picture}
}
%--------------------------------------------
%       o-o-o    3-Punkt-Cluster
%--------------------------------------------
\newsavebox{\kdrei}
\savebox{\kdrei}{
  \begin{picture}(10,10)
    \put( 0, 5){\circle*{2}}
    \put( 0, 5){\line(1,0){5}}
    \put( 5, 5){\circle*{2}}
    \put( 5, 5){\line(1,0){5}}
    \put(10, 5){\circle*{2}}
  \end{picture}
}
%--------------------------------------------
%       o-o    2-Punkt-Cluster
%--------------------------------------------
\newsavebox{\kzwei}
\savebox{\kzwei}{
  \begin{picture}(10,10)
    \put( 2.5, 5){\circle*{2.5}}
    \put( 2.5, 5){\line(1,0){5}}
    \put( 7.5, 5){\circle*{2.5}}
  \end{picture}
}
%--------------------------------------------
%       o    1-Punkt-Cluster
%--------------------------------------------
\newsavebox{\keins}
\savebox{\keins}{
  \begin{picture}(10,10)
    \put( 5, 5){\circle*{2.5}}
  \end{picture}
}
%
%
%
%-------------------------------------------------------------------------
% Definition von Grafikelementen (gross) (vuer CVM)
%-------------------------------------------------------------------------
\unitlength0.6mm
%-------------------------------------------------------------------------
%       o-o-o
%       | | |
%       o-o-o   9-Punkt-Cluster
%       | | |
%       o-o-o
%--------------------------------------------
\newsavebox{\neun}
\savebox{\neun}{
  \begin{picture}(10,10)
    \put( 0, 0){\circle*{2.5}}
    \put( 0, 0){\line(1,0){5}}
    \put( 0, 0){\line(0,1){5}}
    \put( 0, 5){\circle*{2.5}}
    \put( 0, 5){\line(1,0){5}}
    \put( 0, 5){\line(0,1){5}}
    \put( 0,10){\circle*{2.5}}
    \put( 0,10){\line(1,0){5}}
    \put( 5, 0){\circle*{2.5}}
    \put( 5, 0){\line(1,0){5}}
    \put( 5, 0){\line(0,1){5}}
    \put( 5, 5){\circle*{2.5}}
    \put( 5, 5){\line(1,0){5}}
    \put( 5, 5){\line(0,1){5}}
    \put( 5,10){\circle*{2.5}}
    \put( 5,10){\line(1,0){5}}
    \put(10, 0 ){\circle*{2.5}}
    \put(10, 5){\circle*{2.5}}
    \put(10, 0){\line(0,1){5}}
    \put(10,10){\circle*{2.5}}
    \put(10, 5){\line(0,1){5}}
  \end{picture}
}
%--------------------------------------------
%       o-o-o
%       | | |    6-Punkt-Cluster
%       o-o-o   
%--------------------------------------------
\newsavebox{\sechs}
\savebox{\sechs}{
  \begin{picture}(10,10)
    \put( 0, 2.5){\circle*{2.5}}
    \put( 0, 2.5){\line(1,0){5}}
    \put( 0, 2.5){\line(0,1){5}}
    \put( 0, 7.5){\circle*{2.5}}
    \put( 0, 7.5){\line(1,0){5}}
    \put( 5, 2.5){\circle*{2.5}}
    \put( 5, 2.5){\line(1,0){5}}
    \put( 5, 2.5){\line(0,1){5}}
    \put( 5, 7.5){\circle*{2.5}}
    \put( 5, 7.5){\line(1,0){5}}
    \put(10, 2.5){\circle*{2.5}}
    \put(10, 7.5){\circle*{2.5}}
    \put(10, 2.5){\line(0,1){5}}
  \end{picture}
}
%--------------------------------------------
%       o-o
%       | |    4-Punkt-Cluster
%       o-o   
%--------------------------------------------
\newsavebox{\vier}
\savebox{\vier}{
  \begin{picture}(10,10)
    \put( 2.5, 2.5){\circle*{2.5}}
    \put( 2.5, 2.5){\line(1,0){5}}
    \put( 2.5, 2.5){\line(0,1){5}}
    \put( 2.5, 7.5){\circle*{2.5}}
    \put( 2.5, 7.5){\line(1,0){5}}
    \put( 7.5, 2.5){\circle*{2.5}}
    \put( 7.5, 2.5){\line(0,1){5}}
    \put( 7.5, 7.5){\circle*{2.5}}
  \end{picture}
}
%--------------------------------------------
%       o-o-o    3-Punkt-Cluster
%--------------------------------------------
\newsavebox{\drei}
\savebox{\drei}{
  \begin{picture}(10,10)
    \put( 0, 5){\circle*{2.5}}
    \put( 0, 5){\line(1,0){5}}
    \put( 5, 5){\circle*{2.5}}
    \put( 5, 5){\line(1,0){5}}
    \put(10, 5){\circle*{2.5}}
  \end{picture}
}
%--------------------------------------------
%       o-o    2-Punkt-Cluster
%--------------------------------------------
\newsavebox{\zwei}
\savebox{\zwei}{
  \begin{picture}(10,10)
    \put( 2.5, 5){\circle*{2.5}}
    \put( 2.5, 5){\line(1,0){5}}
    \put( 7.5, 5){\circle*{2.5}}
  \end{picture}
}
%--------------------------------------------
%       o    1-Punkt-Cluster
%--------------------------------------------
\newsavebox{\eins}
\savebox{\eins}{
  \begin{picture}(10,10)
    \put( 5, 5){\circle*{2.5}}
  \end{picture}
}
%--------------------------------------------
%            0-Punkt-Cluster
%--------------------------------------------
\newsavebox{\leer}
\savebox{\leer}{
  \begin{picture}(10,10)
    \put( 5, 5){\circle*{0}}
  \end{picture}
}
%--------------------------------------------
%            Pfeil
%--------------------------------------------
\newsavebox{\pfeil}
\savebox{\pfeil}{
  \begin{picture}(5,10)
    \put( 0, 5){\thicklines \vector(1,0){8}}
  \end{picture}
}
%--------------------------------------------
%            leerer Pfeil
%--------------------------------------------
\newsavebox{\pfleer}
\savebox{\pfleer}{
  \begin{picture}(5,10)
    \put( 0, 5){\thicklines \line(1,0){0}}
  \end{picture}
}
%--------------------------------------------
% Definition von Grafikelementen (gross, hohl) (vuer CVM)
%-------------------------------------------------------------------------
\unitlength0.6mm
%-------------------------------------------------------------------------
%       o-o-o
%       | | |
%       o-o-o   9-Punkt-Cluster
%       | | |
%       o-o-o
%--------------------------------------------
\newsavebox{\hneun}
\savebox{\hneun}{
  \begin{picture}(10,10)
    \put( 1.25, 0   ){\line(1,0){2.5}}
    \put( 0   , 1.25){\line(0,1){2.5}}
    \put( 1.25, 5   ){\line(1,0){2.5}}
    \put( 0   , 6.25){\line(0,1){2.5}}
    \put( 1.25,10   ){\line(1,0){2.5}}
    \put( 6.25, 0   ){\line(1,0){2.5}}
    \put( 5   , 1.25){\line(0,1){2.5}}
    \put( 6.25, 5   ){\line(1,0){2.5}}
    \put( 5   , 6.25){\line(0,1){2.5}}
    \put( 6.25,10   ){\line(1,0){2.5}}
    \put(10   , 1.25){\line(0,1){2.5}}
    \put(10   , 6.26){\line(0,1){2.5}}
    \put( 0, 0){\circle{2.5}}
    \put( 0, 5){\circle{2.5}}
    \put( 0,10){\circle{2.5}}
    \put( 5, 0){\circle{2.5}}
    \put( 5, 5){\circle{2.5}}
    \put( 5,10){\circle{2.5}}
    \put(10, 0){\circle{2.5}}
    \put(10, 5){\circle{2.5}}
    \put(10,10){\circle{2.5}}
  \end{picture}
}
%--------------------------------------------
%       o-o-o
%       | | |    6-Punkt-Cluster
%       o-o-o   
%--------------------------------------------
\newsavebox{\hsechs}
\savebox{\hsechs}{
  \begin{picture}(10,10)
    \put( 1.25, 2.5 ){\line(1,0){2.5}}
    \put( 0   , 3.75){\line(0,1){2.5}}
    \put( 1.25, 7.5 ){\line(1,0){2.5}}
    \put( 6.25, 2.5 ){\line(1,0){2.5}}
    \put( 5   , 3.75){\line(0,1){2.5}}
    \put( 6.25, 7.5 ){\line(1,0){2.5}}
    \put(10   , 3.75){\line(0,1){2.5}}
    \put( 0, 2.5){\circle{2.5}}
    \put( 0, 7.5){\circle{2.5}}
    \put( 5, 2.5){\circle{2.5}}
    \put( 5, 7.5){\circle{2.5}}
    \put(10, 2.5){\circle{2.5}}
    \put(10, 7.5){\circle{2.5}}
  \end{picture}
}
%--------------------------------------------
%       o-o
%       | |    4-Punkt-Cluster
%       o-o   
%--------------------------------------------
\newsavebox{\hvier}
\savebox{\hvier}{
  \begin{picture}(10,10)
    \put( 3.75, 2.5 ){\line(1,0){2.5}}
    \put( 2.5 , 3.75){\line(0,1){2.5}}
    \put( 3.75, 7.5 ){\line(1,0){2.5}}
    \put( 7.5 , 3.75){\line(0,1){2.5}}
    \put( 2.5, 2.5){\circle{2.5}}
    \put( 2.5, 7.5){\circle{2.5}}
    \put( 7.5, 2.5){\circle{2.5}}
    \put( 7.5, 7.5){\circle{2.5}}
  \end{picture}
}
%--------------------------------------------
%       o-o-o    3-Punkt-Cluster
%--------------------------------------------
\newsavebox{\hdrei}
\savebox{\hdrei}{
  \begin{picture}(10,10)
    \put( 1.25, 5){\line(1,0){2.5}}
    \put( 6.25, 5){\line(1,0){2.5}}
    \put( 0, 5){\circle{2.5}}
    \put( 5, 5){\circle{2.5}}
    \put(10, 5){\circle{2.5}}
  \end{picture}
}
%--------------------------------------------
%       o-o    2-Punkt-Cluster
%--------------------------------------------
\newsavebox{\hzwei}
\savebox{\hzwei}{
  \begin{picture}(10,10)
    \put( 3.75, 5){\line(1,0){2.5}}
    \put( 2.5, 5){\circle{2.5}}
    \put( 7.5, 5){\circle{2.5}}
  \end{picture}
}
%--------------------------------------------
%       o    1-Punkt-Cluster
%--------------------------------------------
\newsavebox{\heins}
\savebox{\heins}{
  \begin{picture}(10,10)
    \put( 5, 5){\circle{2.5}}
  \end{picture}
}
%--------------------------------------------
%
%
%
%--------------------------------------------
% Definition von Grafikelementen () (vuer CVM)
%-------------------------------------------------------------------------
\unitlength0.8mm
%--------------------------------------------
%       o    1-Punkt-Cluster
%--------------------------------------------
\newsavebox{\einsp}
\savebox{\einsp}{
  \begin{picture}(10,10)
    \put( 5, 5){\circle*{2.5}}
  \end{picture}
}
\newsavebox{\einsm}
\savebox{\einsm}{
  \begin{picture}(10,10)
    \put( 5, 5){\circle{2.5}}
  \end{picture}
}
%--------------------------------------------
%       o-o    2-Punkt-Cluster
%--------------------------------------------
\newsavebox{\zweipp}
\savebox{\zweipp}{
  \begin{picture}(10,10)
    \put( 3.75, 5){\line(1,0){2.5}}
    \put( 2.5, 5){\circle*{2.5}}
    \put( 7.5, 5){\circle*{2.5}}
  \end{picture}
}
\newsavebox{\zweipm}
\savebox{\zweipm}{
  \begin{picture}(10,10)
    \put( 3.75, 5){\line(1,0){2.5}}
    \put( 2.5, 5){\circle*{2.5}}
    \put( 7.5, 5){\circle{2.5}}
  \end{picture}
}
\newsavebox{\zweimm}
\savebox{\zweimm}{
  \begin{picture}(10,10)
    \put( 3.75, 5){\line(1,0){2.5}}
    \put( 2.5, 5){\circle{2.5}}
    \put( 7.5, 5){\circle{2.5}}
  \end{picture}
}
%--------------------------------------------
%       o-o
%       | |    4-Punkt-Cluster
%       o-o   
%--------------------------------------------
\newsavebox{\vierpppp}
\savebox{\vierpppp}{
  \begin{picture}(10,10)
    \put( 3.75, 2.5 ){\line(1,0){2.5}}
    \put( 2.5 , 3.75){\line(0,1){2.5}}
    \put( 3.75, 7.5 ){\line(1,0){2.5}}
    \put( 7.5 , 3.75){\line(0,1){2.5}}
    \put( 2.5, 2.5){\circle*{2.5}}
    \put( 2.5, 7.5){\circle*{2.5}}
    \put( 7.5, 2.5){\circle*{2.5}}
    \put( 7.5, 7.5){\circle*{2.5}}
  \end{picture}
}
\newsavebox{\vierpppm}
\savebox{\vierpppm}{
  \begin{picture}(10,10)
    \put( 3.75, 2.5){\line(1,0){2.5}}
    \put( 2.5 , 3.75){\line(0,1){2.5}}
    \put( 3.75, 7.5 ){\line(1,0){2.5}}
    \put( 7.5 , 3.75){\line(0,1){2.5}}
    \put( 2.5, 2.5){\circle*{2.5}}
    \put( 2.5, 7.5){\circle*{2.5}}
    \put( 7.5, 2.5){\circle{2.5}}
    \put( 7.5, 7.5){\circle*{2.5}}
  \end{picture}
}
\newsavebox{\vierppmm}
\savebox{\vierppmm}{
  \begin{picture}(10,10)
    \put( 3.75, 2.5){\line(1,0){2.5}}
    \put( 2.5 , 3.75){\line(0,1){2.5}}
    \put( 3.75, 7.5 ){\line(1,0){2.5}}
    \put( 7.5 , 3.75){\line(0,1){2.5}}
    \put( 2.5, 2.5){\circle{2.5}}
    \put( 2.5, 7.5){\circle*{2.5}}
    \put( 7.5, 2.5){\circle{2.5}}
    \put( 7.5, 7.5){\circle*{2.5}}
  \end{picture}
}
\newsavebox{\vierpmmp}
\savebox{\vierpmmp}{
  \begin{picture}(10,10)
    \put( 3.75, 2.5){\line(1,0){2.5}}
    \put( 2.5 , 3.75){\line(0,1){2.5}}
    \put( 3.75, 7.5 ){\line(1,0){2.5}}
    \put( 7.5 , 3.75){\line(0,1){2.5}}
    \put( 2.5, 2.5){\circle*{2.5}}
    \put( 2.5, 7.5){\circle*{2.5}}
    \put( 7.5, 2.5){\circle{2.5}}
    \put( 7.5, 7.5){\circle{2.5}}
  \end{picture}
}
\newsavebox{\viermmpp}
\savebox{\viermmpp}{
  \begin{picture}(10,10)
    \put( 3.75, 2.5){\line(1,0){2.5}}
    \put( 2.5 , 3.75){\line(0,1){2.5}}
    \put( 3.75, 7.5 ){\line(1,0){2.5}}
    \put( 7.5 , 3.75){\line(0,1){2.5}}
    \put( 2.5, 2.5){\circle*{2.5}}
    \put( 2.5, 7.5){\circle{2.5}}
    \put( 7.5, 2.5){\circle*{2.5}}
    \put( 7.5, 7.5){\circle{2.5}}
  \end{picture}
}
\newsavebox{\viermppm}
\savebox{\viermppm}{
  \begin{picture}(10,10)
    \put( 3.75, 2.5){\line(1,0){2.5}}
    \put( 2.5 , 3.75){\line(0,1){2.5}}
    \put( 3.75, 7.5 ){\line(1,0){2.5}}
    \put( 7.5 , 3.75){\line(0,1){2.5}}
    \put( 2.5, 2.5){\circle{2.5}}
    \put( 2.5, 7.5){\circle{2.5}}
    \put( 7.5, 2.5){\circle*{2.5}}
    \put( 7.5, 7.5){\circle*{2.5}}
  \end{picture}
}
\newsavebox{\vierpmpm}
\savebox{\vierpmpm}{
  \begin{picture}(10,10)
    \put( 3.75, 2.5 ){\line(1,0){2.5}}
    \put( 2.5 , 3.75){\line(0,1){2.5}}
    \put( 3.75, 7.5 ){\line(1,0){2.5}}
    \put( 7.5 , 3.75){\line(0,1){2.5}}
    \put( 2.5, 2.5){\circle{2.5}}
    \put( 2.5, 7.5){\circle*{2.5}}
    \put( 7.5, 2.5){\circle*{2.5}}
    \put( 7.5, 7.5){\circle{2.5}}
  \end{picture}
}
\newsavebox{\viermpmp}
\savebox{\viermpmp}{
  \begin{picture}(10,10)
    \put( 3.75, 2.5){\line(1,0){2.5}}
    \put( 2.5 , 3.75){\line(0,1){2.5}}
    \put( 3.75, 7.5 ){\line(1,0){2.5}}
    \put( 7.5 , 3.75){\line(0,1){2.5}}
    \put( 2.5, 2.5){\circle*{2.5}}
    \put( 2.5, 7.5){\circle{2.5}}
    \put( 7.5, 2.5){\circle{2.5}}
    \put( 7.5, 7.5){\circle*{2.5}}
  \end{picture}
}
\newsavebox{\vierpmmm}
\savebox{\vierpmmm}{
  \begin{picture}(10,10)
    \put( 3.75, 2.5){\line(1,0){2.5}}
    \put( 2.5 , 3.75){\line(0,1){2.5}}
    \put( 3.75, 7.5 ){\line(1,0){2.5}}
    \put( 7.5 , 3.75){\line(0,1){2.5}}
    \put( 2.5, 2.5){\circle{2.5}}
    \put( 2.5, 7.5){\circle*{2.5}}
    \put( 7.5, 2.5){\circle{2.5}}
    \put( 7.5, 7.5){\circle{2.5}}
  \end{picture}
}
\newsavebox{\viermmmm}
\savebox{\viermmmm}{
  \begin{picture}(10,10)
    \put( 3.75, 2.5){\line(1,0){2.5}}
    \put( 2.5 , 3.75){\line(0,1){2.5}}
    \put( 3.75, 7.5 ){\line(1,0){2.5}}
    \put( 7.5 , 3.75){\line(0,1){2.5}}
    \put( 2.5, 2.5){\circle{2.5}}
    \put( 2.5, 7.5){\circle{2.5}}
    \put( 7.5, 2.5){\circle{2.5}}
    \put( 7.5, 7.5){\circle{2.5}}
  \end{picture}
}
%--------------------------------------------

\eqnnumber
\input{epsf}
\begin{abstract}
On the basis of the one--band $t$--$t'$--Hubbard model a self--consistent
renormalization theory of magnetic short--range order (SRO) in the
paramagnetic phase
is presented combining the four--field slave--boson functional--integral 
scheme with the cluster variational method. Contrary to previous SRO
approaches the SRO is incorporated at the
saddle--point and pair--approximation levels. A detailed numerical evaluation
of the theory is performed at zero temperature, where both the hole-- and
electron--doped cases as well as band--structure effects are studied. The
ground--state phase diagram shows the suppression of magnetic long--range order
in favour of a paramagnetic phase with antiferromagnetic SRO in a wide
doping region. In this phase the uniform static spin susceptibility
increases upon doping up to the transition to the
Pauli paraphase. Comparing the theory with experiments on high--$T_c$ cuprates
a good agreement is found.
\end{abstract}
\parindent0.8cm
{\bf PACS number(s):} {\it 75.10.-b, 71.28.+d, 71.45.-d} 
\thispagestyle{empty}
\newpage
\section{Introduction}
In the theory of strongly correlated itinerant electron systems two
topics are of continued interest, namely (i) the localized--itinerant
complementarity and (ii) the competition between magnetic long--range
order (LRO) and short--range order (SRO). In particular, the concept
of magnetic SRO and its interrelation to itinerant properties was 
investigated in the context of both narrow--band magnetism of
transition metals and their compounds
% \cite{MT78,Mo81,KMP77,PK79,Ha81b,Ka81a} 
[1--6] 
and of
the unconventional magnetic behaviour of high--$T_c$ copper oxides
% \cite{Ka94,Ka91,BSB93,WT90,KM94,TIF95}.
[7--12]. 
In the cuprates, neutron scattering
\cite{RMea93a} and nuclear magnetic resonance experiments \cite{HBSBB90}
reveal pronounced antiferromagnetic spin
correlations within the $\rm CuO_2$ planes which persist even in the
superconducting phase. Moreover, measurements of the spin
susceptibility $\chi(T,\delta)$ in the normal metallic state of $\rm
La_{2-\delta}Sr_\delta CuO_4$ (LSCO) \cite{Toea89,Jo89} show a maximum
in the doping dependence as well as (for $\delta \lapro 0.21$) in the
temperature dependence and give further evidence for strong
SRO effects  at low temperatures, where the SRO decreases with
increasing doping and temperature. Thus the experiments on cuprates
bring out the importance of electron correlations as compared with the
situation in traditional band magnetism, and therefore 
yield a new  challenge for a microscopic theory of SRO in itinerant electron
systems. Such a theory has to provide a self--consistent description of strong
SRO (in the absence of LRO) down to zero temperature. 

The previous theoretical approaches to the problem of SRO in 
transition metals and cuprates are mostly based on
Hubbard--type models
%\cite{MT78,Mo81,PK79,Ha81b,Ka81a,Ka91,BSB93,WT90,KM94,TIF95,HSSJ90}.
[1, 2, 4--6, 8--12, 17]. 
For
example, to explain
the normal--state susceptibility of LSCO in terms of SRO, the
three--band Hubbard model was used \cite{BSB93}. In connection with
the study of SRO some general
problems have gained a renewed interest, such as the stability of various
magnetic LRO phases against paraphases with and without SRO as well as
the question of phase separation 
in strong correlation models \cite{SM94,TIF95}.

From the methodical point of view, the SRO theories for itinerant
systems are preferentially formulated within functional--integral
representations of the one--band Hubbard model using 
the static approximation. Thereby, various
Hubbard--Stratonovich two--field methods, often combined with the
single--site coherent potential approximation (CPA)
\cite{MT78,Mo81,Ha81b,Ka81a,Ka91} and, in the context of
high--$T_c$'s, the scalar four--field slave--boson approach
\cite{KR86,BSB93,TIF95} where employed. Those theories are designed to
describe the formation and ordering of local magnetic moments in an
{\it itinerant} system on a time scale large compared to the electron
hopping time.

In the mode--mode coupling theory of spin fluctuations (SF) by Moriya
{\it et al.} \cite{MT78,Mo81} interpolating between the weakly
magnetic (local SF in $\vec{q}$--space) and local moment (local SF in
real space) limits, the SRO is reflected in the spatial correlation of
thermal SF (with an amplitude increasing with temperature) and is
appreciable even above the Curie temperature. Starting from the local
moment limit and using the two--vector--field
Hubbard--Stratonovich/CPA approach, the SRO is taken into account by
an expansion
in pairwise terms (within the bilinear approximation) around the {\it
single--site} CPA saddle--point. The lack of self--consistency in this
approach may be
justified when the SRO is weak. Note that, at $T=0$, the theory by
Moriya {\it et al.}
\cite{MT78,Mo81} reduces to the Hartree--Fock approximation in both
the weak-- and strong--coupling limits, and the SRO is lost. The
neglect of important correlations at zero temperature may be the
reason for the absence of a maximum in the calculated spin
susceptibility. Contrary, the phenomenological
version of the interpolation theory given in Ref.~\cite{Mo81}
incorporates the SRO  at the saddle point, and in the weakly
magnetic limit the results of the classical approximation to the
self--consistent renormalization theory of SF are recovered. 

In the local--band theory of ferromagnetism \cite{KMP77,PK79}, where
the fluctuations of the local magnetizations with a fixed amplitude
are considered to be local in $\vec{q}$--space, the SRO is described by
an expansion around the Stoner saddle point \cite{PK79} and turns out
to be strong near the Curie temperature. As in Ref.~\cite{MT78}, the
SRO--induced renormalization of the saddle point is disregarded.

The non--self consistency is also inherent in the SRO theories taking the
SF to be local in real space and employing various cluster methods
\cite{Ha81b,Ka81a,Ka91,BSB93}. For example, in the approaches of
Refs.~\cite{Ka81a} and \cite{BSB93} resulting in an effective Ising
free--energy functional, where the exchange energy is evaluated beyond
the bilinear approximation used in Ref.~\cite{MT78} and is nearly 
independent on temperature, the SRO is treated by an expansion around the
single--site CPA saddle--point and by the Bethe--Peierls approximation
\cite{Be35}.

To sum up, all previous microscopic theories based on 
Hubbard--Stratonovich or 
slave--boson/CPA approaches do not take into account 
the SRO self--consistently. Just that, however, is needed when we are
dealing with a {\it strong} SRO at {\it low} temperatures.

Besides the functional--integral techniques, there are only a few
alternative methods concerned with the description of SRO
\cite{WT90,KM94}. For example, to discuss the doping dependence of the
zero--temperature susceptibility $\chi(0,\delta)$ in the one--band
Hubbard model by a semi--phenomenological weak--coupling approach
\cite{KM94}, it was argued that the susceptibility of an
antiferromagnet with only SRO can be approximated by the perpendicular
susceptibility of an antiferromagnet with LRO.

In previous work \cite{TIF95} we have
briefly sketched an 
improved (non--CPA) treatment of SRO in the paraphase of the two--dimensional
(2D) one--band Hubbard model, based on a four--field slave--boson (SB)
functional--integral approach,  by taking into consideration the
feed--back effect of SRO on the saddle point.
In the present paper our self--consistent renormalization theory of
SRO, being valid also at $T=0$, is presented and
discussed in more detail. As compared with Ref.~\cite{TIF95}, the
theory is improved methodically, and several extensions and new
results are given.
For example,  we investigate the
influence of the band structure and of both electron and hole dopings
on the stability of magnetic SRO versus LRO and on the uniform static
spin susceptibility in the paraphase.

The basic idea of our scenario for the occurrence of SRO 
in itinerant electron
systems may be understood as follows. Measuring the
`localization' of the electron spins by the local (not thermally
induced) magnetic moment, with increasing correlation strength,
i.e. with increasing $U/t$ and decreasing doping, the local moment
increases. Concomitantly, local magnetizations develop and an 
effective exchange interaction
will tend to align neighbouring local magnetizations antiferromagnetically,
even in the absence of LRO. Of course, the occurrence of SRO depends on
the doping level, and SRO effects become strongest at $T=0$. That
is, if the
local--moment behaviour is sufficiently pronounced as compared with
the itinerancy, a paraphase with antiferromagnetic SRO becomes
energetically favourable. Let us emphasize that our SRO concept
differs from that described, e.g., in
Refs.~\cite{MT78,Mo81}. 

The paper is organized as follows. In Sec.~2, the action of the SB
functional integral for the partition function of the one--band
$t$--$t'$--Hubbard model on a square lattice (which may describe the essential
features of low--energy charge and spin excitations in the cuprates
\cite{HSSJ90}) is expressed in terms of fluctuating local
magnetizations and internal magnetic fields and transformed
to an effective generalized Ising model with exchange integrals of
arbitrary range. The SRO is taken into account by the
summation over the Ising spins within the cluster variational method (CVM)
\cite{HB55}. In Sec.~3, the resulting effective
free--energy functional is treated in a non--local saddle--point 
approximation,
where the SRO is incorporated, at an equal level of variational
approximation, in a fully self--consistent way. Sec.~4 is devoted to
the calculation of the uniform static spin susceptibility, where the
analytical results on the influence of SRO are given within the
Bethe--Peierls approximation. In Sec.~5, at $T=0$, numerical 
results for the local magnetic moment and magnetization amplitude, the
ground--state phase diagram, and the spin susceptibility as functions
of both the
hole and electron doping level and of the interaction strength are
presented. Thereby, the influence of the band structure, as
parameterized by the nearest-- and next--nearest--neighbour transfer
integrals, is studied. The summary of
our work can be found in Sec.~6.

\section{Effective free--energy functional for the Hubbard model}
\subsection{Slave--boson functional integral}
In the scalar four--field SB representation~\cite{KR86}
the Fock space at each site is enlarged by introducing the Bose
fields $e_i$,  $p_{i\sigma}$ and $d_i$ describing projection
operators onto empty, singly and doubly occupied states,
respectively, and the Hubbard model is expressed as 
\begin{equation}
\cH=\sum_{ij\sigma}t_{ij}^{ } z_{i\sigma}^{\dagger}
f_{i\sigma}^{\dagger}
f_{j\sigma}^{ } z_{j\sigma}^{ } +U\sum_i d_i^\dagger d_i^{ } 
- h \sum_{i\sigma} \sigma f_{i\sigma}^{\dagger}f_{i\sigma}^{} \;,
\label{ho}
\end{equation}
where $t_{ij}$ are the transfer integrals between
nearest ($-t$) and next--nearest  ($-t^{\prime}$) neighbours,  
$h$ denotes the uniform external magnetic field, and 
\begin{equation}
\zis=(1-\diddi-\pisdpis)^{-\frac{1}{2}}
     (\eid\pis+\pimsd \di)
     (1-\eidei-\pimsdpims)^{-\frac{1}{2}}\,.
\label{zfa}
\end{equation}
To exclude unphysical states in the extended Fock space, the
local constraints
\begin{equation}
\eidei+\diddi+\sum_{\sigma} \pisdpis =1\;,
\label{co1}
\end{equation}
\begin{equation}
\pisdpis+\diddi =\fisdfis
\label{co2}
\end{equation}
have to be fulfilled. Expressing the partition function $\cZ$ by
a coherent--state functional integral over complex Bose 
($\{b^{(*)}\};\;b=e,\,p_{\sigma},\,d$) and
pseudofermionic Grassmann ($\{f,\bar{f}\}$) fields, 
the local constraints are enforced by the 
time--independent Lagrange multipliers $\lambda_{i}^{(1)}$ and
$\lambda_{i\sigma}^{(2)}$. Integrating out the pseudofermions we have
\begin{eqnarray}
\cZ&=&\int [\cD d_{ }^{(*)}][\cD e_{ }^{(*)}]
[\cD p_\sigma^{(*)}][d\lambda_{
}^{(1)}][d\lambda_{\sigma}^{(2)}]
\exp \Big\{  -\int_0^\beta d\tau \cL (\tau)\Big\}
\label{z1}\;,\\[0.2cm]
\cL (\tau) &=& \sum_i \Big[ 
\eic (\partial \tau + \lbi)\ei 
+\sum_\sigma \pisc (\partial \tau + \lbi - \lbis )\pis
\nonumber \\
& &  \hspace*{2cm}
 +\dic \Big(\partial \tau + \lbi - \sum_\sigma \lbis+U\Big)\di  
-\lbi \Big]- \mbox{Tr ln} \left[- G_{ij\sigma}^{-1}(\tau)\right]
\label{la}
\end{eqnarray}
with
\begin{equation}
G_{ij\sigma}^{-1}(\tau) = 
(-\partial_{\tau} +\mu - \lambda_{i\sigma}^{(2)}+ \sigma h)\,\delta_{ij}^{ } - 
z_{i\sigma}^{*}\,z_{j\sigma}^{ }t_{ij}^{}\;.
\label{gpi}
\end{equation}
To remove the non--diagonal fluctuations in the transfer term of the
inverse Green propagator~(\ref{gpi}), we employ the Shiba
transformation~\cite{Sh71a} 
\begin{equation}
G_{ij\sigma}^{ } \to 
\hG = \sum_{lm} \hat{z}_{il\sigma}^{*} G_{lm\sigma}^{ }
\hat{z}_{mj\sigma}^{ } \,,
\label{shiba}
\end{equation}
where $\hat{z}_{ij\sigma}^{ } = \zis \delta_{ij}$.
Then, in the frequency representation of the Lagrangian (\ref{la}),
under the trace operation $G_{ij\sigma}^{-1}(\omega)$ 
can be replaced exactly by
\begin{equation}
\ihG(\omega+\mu)=\frac{\omega+\mu-\lbis+\sigma h}{\qis} \,\delta_{ij}
-t_{ij} 
\label{ghat}
\end{equation}
with $q_{i\sigma}=|z_{i\sigma}|^2$.
By a local  $U(1)^{\otimes 3}$ gauge transformation the phases of the
Bose fields $e_i$ and $p_{i\sigma}$ are removed (radial gauge), 
and the Lagrange multipliers become time--dependent fields.
Applying the static approximation for the bosons
($p_{i\sigma}^{ }\,,d_i^{ }\,,\lambda_{i\sigma}^{(2)}$, where the 
saddle--point approximation for $\lambda_i^{(1)}$ is used to eliminate the
integrals over the $e_i$ fields) and adopting the transformed 
fields~\cite{Ha89} 
\begin{eqnarray}
\label{Hase1}
 && \mi = \sum_{\sigma}\sigma p_{i\sigma}^2 \;, 
 \;\;\;\;\;\;\;\;\,
 \xii = -{\mbox{$\frac{1}{2}$}}\sum_\sigma \sigma \lambda_{i\sigma}^{(2)} \;,
\\
\label{Hase2}
 && \ni = \sum_{\sigma} p_{i\sigma}^2 + 2 \dicdi \;, 
 \;\;\;\;\nui =
{\mbox{$\frac{1}{2}$}}\sum_\sigma\lambda_{i\sigma}^{(2)} \;,
\end{eqnarray}
the partition function is obtained as
\begin{eqnarray}
\label{z2}
\cZ &=& \int [\cD d][\cD d^{*}][\cD n][\cD \nu][\cD m][\cD
\xi]\exp{\left\{- \beta {\mit \Psi}(\{d,d_{ }^*,n,\nu,m,\xi\})
\right\}}\;,
\\[0.2cm]
{\mit \Psi}&=& \sum_i\left(U  \dicdi
- n_i \nu_i + m_i \xi_i\right) +\mbox{$\frac{1}{\pi}$}\int d\omega f(\omega 
-\mu) \,\mbox{Im}\;\mbox{Tr} 
\ln{\left[-\hat{G}_{ij\sigma}^{-1}(\omega)\right]}\;. 
\label{PSIO}
\end{eqnarray}
In eqs.~(\ref{Hase1}) and~(\ref{Hase2}), $m_i^{ }$ and $n_i$ are the bosonic 
representations of the local magnetization and particle number, respectively, 
defined analogously to its fermionic counterparts $m_i^f=\sum_\sigma
\sigma f_{i\sigma}^\dagger f_{i\sigma}^{}$ and
$n_i^f=\sum_\sigma f_{i\sigma}^\dagger f_{i\sigma}^{}$. 
Since $\xi_i^{ }$ couples to $m_i$ as a magnetic field 
we denote $\xi_i^{ }$ by `internal magnetic field'.
\subsection{Transformation to a generalized Ising model}
To incorporate the SRO beyond the {\it uniform} paramagnetic (PM)
saddle point, we perform an expansion in terms of the local
perturbation $\Vis \delta_{ij}^{ }=-\ihG+\ihGo$, 
where
\begin{equation}
\hGo(\omega)=\frac{\qos}{N}\sum_{\vec{k}}
\frac{ \mbox{e}^{i\vec{k}(\vec{R}_i-\vec{R}_j)}}{\omega-\Eks}
\label{pmgp}
\end{equation}
is the PM saddle--point propagator with
\begin{equation}
\Eks=\lbos -\sigma h + q_{\sigma}^o \ek\;,
\label{qten}
\end{equation}
$\lbos=\nu_{ }^o - \sigma \xi_{ }^o$, and the 2D tight--binding band dispersion
\begin{equation}
\ek=-2t(\cos \kx+ \cos \ky)-4 \ts \cos \kx \cos \ky\;.
\label{badi}
\end{equation}
By the diagonal part of the Green propagator~(\ref{pmgp}), the
perturbation reads
\begin{equation}
\Vis(\omega)= \left(1-\frac{\qos}{\qis}\right)
\ihGoii(\omega)+\frac{\lbis-\lbos}{\qis}\;,
\label{stoe}
\end{equation}
and we can split up the fermionic part of~(\ref{PSIO}) as
\begin{eqnarray}
\label{Tr_I}
\mbox{Tr}~\ln{\left[-\ihG(\omega)\right]} &=&
\mbox{Tr}~\ln{\left[-\ihGo(\omega)\right]} \nonumber \\
& & \hspace*{-0.3cm}+ \mbox{Tr}~\ln{\left[(1-\hGoii(\omega)\Vis(\omega)
)\delta_{ij}\right]} \nonumber \\
& & \hspace*{-0.3cm}+
\mbox{Tr}~\ln{\left[\delta_{ij}-\hGoij(\omega)\Tjs(\omega)\right]}\,.
\end{eqnarray}  
Here, $\Tis=\Vis\,[1-\hGoii\Vis]^{-1}$ is the scattering
matrix, and $\hGoij$ denotes the nondiagonal part of~(\ref{pmgp}).
The second term on the r.h.s. of (\ref{Tr_I}) yields a single--site 
fluctuation contribution to the functional (\ref{PSIO}), whereas the third 
term describes the coupling between the
fluctuations at all sites and is responsible for SRO effects.

Treating the fluctuations of the local
magnetizations $m_i^{ }$ and the internal magnetic fields $\xi_i^{ }$ on
an equal footing, we express those fields by their amplitude and direction
according to 
\begin{equation}
\label{locMag}
m_i^{ } = \mb_i^{ }\si \;,\;\;
\xi_i^{ } = \xib_i^{ }\si \;,\;\;\si = \pm \;.
\end{equation}
Furthermore, we make the ansatz
\begin{equation}
\label{Ansatz}
b_i \to b_{s_i}\;\mbox{with}\;\; b=\mb, \xib ,n , \nu, d, d_{ }^{*}\;,
\end{equation}
which becomes relevant for the calculation of the uniform static spin 
susceptibility (Sec.~4); for $h=0$,
we have $b_{s_i}=b$. 
In Appendix~A the resulting partition function is transformed to that
of an effective generalized Ising model along the lines indicated  
by Kakehashi~\cite{Ka81a}. In the pair approximation, where all terms
of~(\ref{Tr_I}) involving more than two sites are neglected, we
obtain
\begin{equation}
\label{PFUNK}
\cZ = \sum_{\{\si\}} \int [\cD d_{\alpha}] [\cD d_{\alpha}^*] [\cD n_{\alpha}][\cD \nu_{\alpha}]
[\cD \bar{m}_{\alpha}][\cD \bar{\xi}_{\alpha}]\exp
\{-\beta {\mit \Psi}(b_{\alpha},\{\si\})\}\;
\end{equation}
with 
\begin{eqnarray}
\label{PsiIs1}
{\mit \Psi}(b_{\alpha},\{\si\}) &=& \bPsi(b_{\alpha}) +
{\mit \Psi}_{ }^{\rm (pair)}(b_{\alpha},\{\si\})\,,\\[0.2cm]
{\mit \Psi}_{ }^{\rm (pair)}(b_{\alpha},\{\si\}) 
&=& -\bar{h}(b_{\alpha}) \sum_i \si 
- \sum_{n} \bar{J}_{n}^{ }(b_{\alpha})
 \sum_{(ij)_n^{}} \si \sj\,.\label{PsiIs2}
\end{eqnarray} 
Here, $\alpha=\pm 1$ and the index $n$ denotes 
the $n$--th neighbour shell around a given
lattice point $i$, where $(ij)_n$ sums over the $z_n$ sites $j $ 
in the $n$--th shell. The expressions for $\bar{h}$ and 
the effective Ising--exchange integrals $\bar{J}_n$ are given in 
Appendix A. Let us emphasize that those quantities 
are complicated functions of all SB fields and have to be determined 
self--consistently at each interaction strength $U$ and hole doping 
$\delta=1-n$.

Furthermore, remark that $\bar{J}_n$ has to be contrasted from the
effective Ising--exchange energy $\tilde{J}$ occurring in previous
Hubbard--Stratonovich/CPA
approaches~\cite{Ka81a,FKK84}, where $\tilde{J}$ couples 
thermally induced local moments.
\subsection{Cluster variational method}
To perform the $s_i$--sum in the partition function~(\ref{PFUNK}) of the
effective Ising model~(\ref{PsiIs2}{}), we restrict ourselves to the square
lattice and employ the CVM in the version given by Hijmans and de
Boer~\cite{HB55}. In this method the calculation of the free energy in
the $(w)$--cluster approximation starts from the choice of a basic
cluster $(w)$ and the series of subclusters $(v)$ generated by
overlapping two basic clusters in all possible ways, as illustrated in
Fig.~1 [$(w)=(f),\;(v)=(e),\ldots, (a)$].
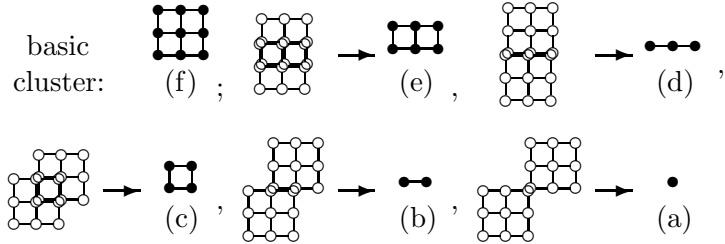
\begin{figure}[ht]
  \unitlength0.6mm
\centerline{
\begin{picture}(160,60)
\put(  5  ,40  ){\mbox{basic}}
\put(  2  ,32  ){\mbox{cluster:}}
\put( 32  ,40  ){\mbox{\usebox{\neun}}}
\put( 35.5,32  ){\mbox{(f)}}
\put( 47  ,31  ){\mbox{;}}
\put( 54.5,32.5){\mbox{\usebox{\hneun}}}
\put( 55  ,38  ){\mbox{\usebox{\hneun}}}
\put( 72  ,35  ){\mbox{\usebox{\pfeil}}}
\put( 84  ,39  ){\mbox{\usebox{\sechs}}}
\put( 87.5,32  ){\mbox{(e)}}
\put( 99 ,31  ){\mbox{,}}
\put(109  ,30  ){\mbox{\usebox{\hneun}}}
\put(109.5,40.5){\mbox{\usebox{\hneun}}}
\put(129  ,35  ){\mbox{\usebox{\pfeil}}}
\put(141  ,37  ){\mbox{\usebox{\drei}}}
\put(144.5,32  ){\mbox{(d)}}
\put(158  ,36  ){\mbox{,}}
\put(  0  , 2.5){\mbox{\usebox{\hneun}}}
\put(  5.5, 8  ){\mbox{\usebox{\hneun}}}
\put( 20  , 5  ){\mbox{\usebox{\pfeil}}}
\put( 32  , 8  ){\mbox{\usebox{\vier}}}
\put( 35.5, 2  ){\mbox{(c)}}
\put( 47  , 6  ){\mbox{,}}
\put( 52  , 0  ){\mbox{\usebox{\hneun}}}
\put( 57.5,10.5){\mbox{\usebox{\hneun}}}
\put( 72  , 5  ){\mbox{\usebox{\pfeil}}}
\put( 84  , 7  ){\mbox{\usebox{\zwei}}}
\put( 87.5, 2  ){\mbox{(b)}}
\put( 99  , 6  ){\mbox{,}}
\put(104  , 0  ){\mbox{\usebox{\hneun}}}
\put(114.5,10.5){\mbox{\usebox{\hneun}}}
\put(129  , 5  ){\mbox{\usebox{\pfeil}}}
\put(141  , 7  ){\mbox{\usebox{\eins}}}
\put(144.5, 2  ){\mbox{(a)}}
\end{picture}}
  \caption{Genesis of the series of subclusters $(e)$ to $(a)$ from the
     basic cluster $(f)$ in the square lattice (see text).}
\end{figure} 
In the same way, the next
generation of overlap figures can be obtained from each of the
subclusters $(v)$. Note that in each generation there occur
`essential' overlap figures [e.g., $(e)$ in Fig.~1] which do not
appear as overlap figures in the subsequent generations.

Following Ref.~\cite{HB55}, the free energy of the {\it N}--site system in the  
$(w)$--cluster approximation is obtained as
\begin{equation}
\cF_{w}(p_{\kappa}^{(w)})=\sum_{v=a}^{w} y_{w}^{(v)} 
\cF^{(v)}(p_{\kappa^{\prime}}^{(v)})\;,
\label{fhb1}
\end{equation}
where
\begin{equation}
\cF_{ }^{(v)}(p_{\kappa}^{(v)})=N
\sum_{\kappa\in {(v)}}\lambda_{\kappa}^{(v)} p_{\kappa}^{(v)}
\left(\varepsilon_{\kappa}^{(v)}
      +\frac{1}{\beta}\ln{p_{\kappa}^{(v)}}\right)
\label{fhb2}
\end{equation}
is the free energy of an assembly of $N$ independent clusters of type
$(v)$. Here, $\kappa$ labels the topologically inequivalent spin
configurations of the cluster $(v)$ with energy
$\varepsilon_{\kappa}^{(v)}$ and the multiplicity
$\lambda_{\kappa}^{(v)}$
(number of equivalent configurations of type $\kappa$),
where $\varepsilon_{\kappa}^{(v)}$ and $\lambda_{\kappa}^{(v)}$ are
calculated for an {\it isolated} 
cluster $(v)$. The distribution number $p_{\kappa}^{(v)}$ is the number of 
$(v)$--clusters with fixed configuration $\kappa$
contained in a given configuration of the
$N$--site lattice divided by the total number of $(v)$--clusters in
the lattice, where the normalization condition 
\begin{equation}
\sum_{\kappa\in(v)} \lambda_{\kappa}^{(v)} p_{\kappa}^{(v)} =1
\label{normcond}
\end{equation}
has to be fulfilled. Of course, the distribution numbers 
$p_{\kappa}^{(v)}$ for the subcluster $(v)$ are determined by the
distribution numbers $p_{\kappa}^{(w)}$ for the choosen basic 
cluster $(w)$, i.e. 
\begin{equation}
p_{\kappa}^{(v)}=p_{\kappa}^{(v)}[p_{\kappa'}^{(w)}]
\;,\;\;\;\forall\;(v)\subset (w)\,.
\label{konrel}
\end{equation} 
Since in the cluster construction of the total free energy in terms
of the  `microvariables' $p_{\kappa}^{(v)}[p_{\kappa'}^{(w)}]$ 
described so far the different subclusters $(v)$ are counted several times, 
in~(\ref{fhb1}) the geometrical factors $y_w^{(v)}$ were introduced to
compensate this overcounting and to give the correct number of
$(v)$--clusters contributing to the free energy. 
In Appendix B the consistency relations~(\ref{konrel}) 
as well as the  linear system of equations for $y_w^{(v)}$
are given in a more explicit form for the most simplest basic clusters
$(a)$, $(b)$, and $(c)$. 

Finally, in the spirit of the CVM, 
the free energy has to be minimized with respect to the 
`macrovariables' $p_{\kappa}^{(w)}$. Since the 
constraint~(\ref{normcond}) also holds for the basic cluster
$(w)$, we can reduce the variational degrees of freedom by one 
introducing, instead of the $p_{\kappa}^{(w)}$, 
a new set of independent variables $x_\iota$.
Accordingly, the CVM equations read 
\begin{equation}
\label{CVMGl}
\frac{\partial \cF_{w}(p_{\kappa}^{(w)}(x_{\iota}))}{\partial x
_{\iota}}=0\,.
\end{equation}
It is worth emphasizing that alternatively all the CVM variables 
$x_\iota$ contained in $ \cF_{w}$ can be expressed 
by the complete set of $M$--point ($M=1,\ldots,N^{(w)}$)
correlation functions $K$
belonging to the basic cluster $(w)$~\cite{Mo57}.

Applying this method to our spin model~(\ref{PsiIs2}) we obtain 
the effective free--energy functional in the $(w)$--CVM approximation
(c.f. Appendix~B),
\begin{eqnarray}
\label{effBA}
{\mit \Psi}^{\rm (pair)}_{w}(b_{\alpha},x_\iota) &=&-N \biggl[
\Kom(x_\iota)\, \bh(b_{\alpha})
+\sum_{n}\frac{z_{n}}{2} \Knm(x_\iota) \bJn(b_{\alpha}) \biggr] 
\nonumber \\
& &+N \sum_{v=a}^{w}y_{w}^{(v)}\sum_{\kappa \in (v)} \lambda_{\kappa}^{(v)}\,
p_{\kappa}^{(v)}(x_\iota)\ln p_{\kappa}^{(v)}(x_\iota)\,,
\end{eqnarray}
where the internal energy part is given in terms of the expectation value 
\begin{equation}
\Kom :=\langle\si\rangle =-\frac{1}{N}\frac{\partial \cF_w}{\partial \bar{h}}
\end{equation}
and of all pair--correlation functions in the basic cluster,
\begin{equation}
\label{korrf}
\Knm :=\langle\si\sj\rangle_n^{}=
-\frac{2}{z_{n}N}\frac{\partial \cF_w}{\partial \bar{J}_n} 
\,.
\end{equation}
In~(\ref{effBA}) the coupling to higher, e.g., 
3--point or 4--point correlation functions results from the entropy term.

\section{Saddle point with short--range order}
\setcounter{equation}{0}
Now we calculate the free energy per site
\begin{equation}
f(n,h,T)=\left.\left[\frac{1}{\beta N} \cS +\mu n\right]\right|_{\rm SP}\;,
\label{freen}
\end{equation}
where 
\begin{equation}
\cS= \beta \left[ \bPsi(b_{\alpha})
+{\mit \Psi}^{\rm (pair)}_{w}(b_{\alpha},x_{\iota}) \right]\;
\label{effact}
\end{equation}
is the effective bosonic action,
adopting the saddle--point and $(w)$--cluster variational 
approximations for all Bose fields $b_{\alpha}$ and variational parameters
$x_{\iota}$, respectively. Taking into account
the dependences on $b_{\alpha}$ of $\bPsi$, $\bh$ and $\bJn$
(cf. Appendix A), from 
$\partial \cS / \partial b_{\alpha}=0$ we get
\begin{eqnarray}
\!\!\!&&\!\!\!\!\!\!
\mbox{$\frac{1}{2}$}\sum_{\alpha'}\left\{\left(1+\alpha' \Kom\right)
\left[\frac{\partial}{\partial b_{\alpha}}(U \dcda -\na \nua +\mba \xiba)
\right.
+\sum_{\sigma}
\frac{\partial {\mit \Phi}_{\alpha \sigma}^{ }}{\partial b_{\alpha}}
\right] \nonumber \\
&&\;\;+\sum_{n\subseteq (\!w\!)\atop \sigma}\frac{z_{n}}{4}\left[
\left(1+2\alpha' \Kom+\Knm\right)
\frac{\partial {\mit \Phi}_{n,\alpha' \alpha' \sigma}^{ }}
{\partial b_{\alpha}} + \left.\left(1-\Knm\right)
\frac{\partial {\mit \Phi}_{n,-\alpha' \alpha' \sigma}^{ }}
{\partial b_{\alpha}}
\right]\right\}=0\;.
\end{eqnarray}
This coupled system of self--consistency equations determining
the saddle point with SRO can be cast to the form
\begin{eqnarray}
\label{SROSattel1}
\mba = \alpha \sum_\sigma \sigma n_{\alpha\sigma}^{f} &,&\quad
\xiba = \displaystyle -\sum_\sigma Q_{\alpha\sigma}^{f} 
\frac{\partial q_{\alpha\sigma}^{ }}{\partial \mba}
\;, \\
\label{SROSattel2}
\na = \sum_\sigma n_{\alpha\sigma}^{f} \;\;\;\;\; &,&\quad
\nua = \sum_\sigma Q_{\alpha \sigma}^{f} 
\frac{\partial q_{\alpha\sigma}^{ }}{\partial \na}\;, 
\end{eqnarray}
\begin{equation}
\label{SROSattel3}
U = -\sum_\sigma Q_{\alpha \sigma}^{f} 
\frac{\partial q_{\alpha\sigma}^{ }}{\partial d_\alpha^2}\;.
\end{equation}
Here, $n_{\alpha \sigma}^{f}$ is the pseudofermionic
expression for the particle density given by
\begin{eqnarray}
\label{NASF}
n_{\alpha \sigma}^{f} \!\!\!&=&\!\!\!  \frac{1}{1+\alpha \Kom} \sum_{\alpha'}
\left\{ \left(1+\alpha' \Kom \right)
\frac{\partial{\mit \Phi}_{\alpha'\sigma}}
{\partial \lbas} \nonumber\right. \\
&&
+\sum_{n}\frac{z_n}{4}\left[
\left(1+2\alpha'\Kom+\Knm\right)
\frac{\partial{\mit \Phi}_{n,\alpha'\alpha'\sigma}}
{\partial \lbas} +\left.\left(1-\Knm \right)
\frac{\partial {\mit \Phi}_{n,{-\alpha'}\alpha'\sigma}}
        {\partial \lbas}\right]\right\}\;,
\end{eqnarray}
where $\lambda_{\alpha \sigma}^{(2)}=\nua - \alpha \sigma \xiba$. 
In eqs.~(\ref{SROSattel1}) to (\ref{SROSattel3}),
$Q_{\alpha \sigma}^{f}$ is given by  
(\ref{NASF}) with $\partial \lambda_{\alpha \sigma}^{(2)}$ replaced
by $\partial q_{\alpha\sigma}$. Finally, the chemical
potential is determined from the number condition
\begin{equation}
n=1-\delta=\sum_{\alpha\sigma} p_{\alpha}^{(a)}\,
n_{\alpha \sigma}^{f}
\end{equation}
with
\begin{equation}
\label{propA}
p_{\alpha}^{(a)}=
\frac{1}{2}(1+\alpha x_{1})\,. 
\end{equation}
At the saddle point, we have 
$n_{\alpha \sigma}^{f}=n_{\alpha \sigma}=(\na + \alpha
\sigma \mba)/2$, where $n_{\alpha \sigma}^{ }$ is the bosonic
expression.
For vanishing local magnetization, $\mba=0$, we have $V_{\alpha\sigma}=0$
and (by (\ref{JIsing}) and (\ref{intPaar})) $\bar{J}_n=0$ so that there is no
SRO, and the PM saddle point (with the external field $h$) is recovered from 
(\ref{SROSattel1}) to (\ref{SROSattel3}). 
Accordingly, in the $h=0$ limit ($\bar{m}_{\alpha}=\bar{m}$; i.e., the
integrals ${\mit \Phi}$ in (\ref{NASF}) only depend on
$\eta=\alpha\sigma =\pm$) we obtain two possible 
paramagnetic phases: (i) the paraphase without SRO (PM) and (ii) the 
paraphase with magnetic SRO (SRO--PM), i.e., \begin{equation}
\label{defphasen}
\begin{array}{c@{\;:\;\;} l@{\;,\;\;} l@{\;;\;\;} l}
\mbox{PM} & 
\langle s_i^{ } \rangle  = 0 &
\langle s_i^{ } s_j^{ }  \rangle = 0 &
\mb = 0 \;, \\
\mbox{SRO--PM}& 
\langle s_i^{ } \rangle  = 0 &
\langle s_i^{ } s_j^{ }  \rangle \neq 0 &
\mb > 0 \;, 
\end{array}
\end{equation}
where $i$ and $j$ are nearest--neighbour sites. 

Let us stress that, at $T=0$, the SRO--PM phase with antiferromagnetic
correlations ($\langle s_i^{ } s_j^{ }  \rangle < 0 $; $\bJone<0$)  
must be distinguished from the phase 
with antiferromagnetic LRO (denoted by AFM) having a
finite sublattice magnetization,
$m_A^{ }=p_{A\uparrow}^{2}-p_{A\downarrow}^{2}=-m_B^{ }$,
which is determined from the A--B saddle--point solution~\cite{DFB92} 
and differs
from the SRO amplitude $\bar{m}$; in particular, we 
found $\mb \neq 0$ in parameter regions,
where $m_A=0$ (see below, Fig.~2).

\section{Spin susceptibility}
\setcounter{equation}{0}
The uniform static spin susceptibility 
$\chi=\lim_{h\to 0} \frac{dm}{d h}$, where 
$m=\sum_\alpha p_\alpha^{(a)} m_\alpha$ with 
$m_{\alpha}^{ }=\alpha \mba$ is the averaged magnetization,
has to be calculated from
\begin{equation}
\label{chidef}
{\mit \chi}=\lim_{h\to 0}\,\sum_\alpha 
\left( p_\alpha^{(a)} \frac{d\,m_\alpha}{d\,h}
+m_\alpha \frac{d\,p_\alpha^{(a)}}{d\,h}\right)\;,
\end{equation}
using the solutions of the saddle--point and cluster variational
equations for $\mba$ and $p_\alpha^{(a)}(x_1)$, respectively. 
The first term in~(\ref{chidef}) describes the change of the
magnetization amplitude with the applied magnetic field and gives
mainly the `itinerant' contribution to $\chi$. The second term
describes directional fluctuations of the magnetizations and is
called the `local' contribution being finite only in the SRO--PM
phase. However, note that the `itinerant' and `local' properties are
interrelated and determine {\it both} contributions to the spin
susceptibility.

Now we exploit the symmetry relations in the paraphase
$\left.\frac{d\,\bar{m}_+}{d\,h}\right|_{h=0}
\equiv\bar{m}_{\scriptscriptstyle +_{\scriptstyle h}}^{ }
= - \bar{m}_{\scriptscriptstyle -_{\scriptstyle h}}^{ }$ and
$\left.\frac{d\,\bar{\xi}_+}{d\,h}\right|_{h=0}
\equiv\bar{\xi}_{\scriptscriptstyle +_{\scriptstyle h}}^{ }
= - \bar{\xi}_{\scriptscriptstyle -_{\scriptstyle h}}^{ }\,.$
Taking into consideration the full dependence on the external field
$h$ of the SB fields [$\bar{m}_+,\,\bar{\xi}_+,\,m^o,\,\xi^o$] 
and variational parameters [$x_\iota$], in the limit $h\to 0$ 
the uniform static spin susceptibility can 
be obtained from the solution of the following linear system of equations 
\begin{eqnarray}
\label{chigs1}
\sum_\iota \cS_{x_\iota y}^{ } x_{\scriptscriptstyle \iota_{\scriptstyle h}}
+\cS_{\bar{m}_{[+,-]}^{ } y}^{ } 
\bar{m}_{\scriptscriptstyle +_{\scriptstyle h}}^{ } 
+\cS_{\bar{\xi}_{[+,-]}^{ } y}^{ } 
\bar{\xi}_{\scriptscriptstyle +_{\scriptstyle h}}^{ } 
+\cS_{m_{ }^o y}^{ } m_h^o
+\cS_{\xi_{ }^o y}^{ } \xi_{h}^o &=&
-\cS_{h y}^{ } \;,\\ 
\label{chigs2}
\cS_{m_{ }^o y^o}^{o} m_h^o +\cS_{\xi_{ }^o y^o}^{o} \xi_h^o 
&=& -\cS_{hy^o}^{o}\;,
\end{eqnarray}
where 
%\begin{equation}
$X_y^{ } = \left.\frac{\partial X}{\partial y}\right|_{h=0} \;,\;\;
X_{x_{[+,-]}^{ }}^{ } =
\left.\left( \frac{\partial }{\partial x_+^{ }} 
- \frac{\partial }{\partial x_-^{ }} \right) X\right|_{h=0}\,,\;\;\;
X_{xy}^{ } = \left.\frac{\partial^2 X}{\partial x \partial y}\right|_{h=0}$
%\label{defabl}
%\end{equation}
with $y=\bar{m}_+^{ },\bar{\xi}_+^{ },x_\iota$ and  $y^o=m_{ }^o,
\xi_{ }^o$. 

In the paraphase without SRO, from~(\ref{chigs2}) we find 
\begin{equation}
\label{chio2}
\xi_h^o = \frac{1+\frac{1}{2}B{\mit \chi}_1^o}
{1+A \, {\mit \chi}_0^o + B \,{\mit\chi}_1^o + \frac{1}{4}B_{ }^2
({\mit\chi}_1^{o2}-{\mit\chi}_0^o{\mit\chi}_2^o)}-1\,,
\end{equation}
and the suceptibility is 
\begin{equation}
\label{chio1}
\chi^o:=m_h^o = \frac{{\mit \chi}_0^o}
{1+A \, {\mit \chi}_0^o + B \,{\mit\chi}_1^o + \frac{1}{4}B_{ }^2
({\mit\chi}_1^{o2}-{\mit\chi}_0^o{\mit\chi}_2^o)}
\end{equation}
with 
\begin{equation}
A = \sum_\sigma \frac{ \partial^2 q_\sigma^o}{\partial m^{o2}}\;,\;\;
B=\frac{1}{2}\sum_\sigma \sigma 
\frac{ \partial  q_\sigma^o}{\partial m^{o}}\,,
\label{coeff}
\end{equation}
and the higher--order Lindhard functions
\begin{equation}
{\mit \chi}_l^o= -\frac{1}{N} \sum_{\vec{k}\sigma} 
(2\varepsilon_{\vec{k}})^l 
\frac{\partial f(E_{\vec{k}\sigma}^o-\mu)}
{\partial E_{\vec{k}\sigma}^o}\,.
\label{chiol}
\end{equation}
Let us point out that our result for the spin susceptibility $\chi^o$
agrees with the static and uniform limit of the dynamic spin
susceptibility derived, within the spin--rotation--invariant 
SB scheme~\cite{LWH89}, from
the Gaussian fluctuation matrix at the PM saddle point~\cite{LSW91}.
Contrary, the uniform static spin susceptibility 
given in Ref.~\cite{La90} within the scalar four--field SB approach 
disagrees with the result of Li {\it et al.}~\cite{LSW91} away from
half--filling. 

If we include magnetic SRO effects at the Bethe--Peierls level
of approximation ($\iota=1,2$; cf. Appendix B), from the
evaluation of~(\ref{chigs1}) together with (\ref{chio2})--(\ref{chiol}) 
we finally obtain the spin susceptibility as
\begin{equation}
{\mit \chi} = \frac{\mit \Theta_{\rm I}^{ } \, \chi_{\rm I}^{ } 
+ \chi_{\rm II}^{ }}
{\mit \Theta_{\rm I}^{ }+ \mit \Theta_{\rm II}^{ }}\;,
\label{chisro}
\end{equation}
where
\begin{eqnarray}
\label{chi1}
{\mit \chi_{\rm I}^{ }} & = & {\mit\Delta}
\left[{\mit \Xi}\, \cS_{\bar{m}_{[+,-]}^{ }\bar{\xi}_+^{ }}^{ }
-{\mit \Lambda} \cS_{\bar{\xi}_{[+,-]}^{ }\bar{\xi}_+^{ }}^{ }\right]
\,,\\[0.2cm]
\label{chi2}
{\mit \chi_{\rm II}^{ }} & = & \beta \bar{m}_+^{ } {\mit \Gamma} +
\beta {\mit\Delta}
\left\{ 
\left[ \bar{m}_+^{ } \left({\mit \Xi}\,\bar{h}_{\bar{m}_{[+,-]}^{ }}^{ }
+ {\mit \Lambda}\,\bar{h}_{\bar{\xi}_{[+,-]}^{ }}^{ }\right)
- {\mit \Gamma} \cS_{x_{1}^{ }\bar{\xi}_+^{ }}^{ }\right]
\cS_{\bar{m}_{[+,-]}^{ }\bar{\xi}_+^{ }} 
\right. \nonumber \\
&& \hspace{1.2cm} 
- \bar{h}_{\bar{\xi}_{[+,-]}^{ }}^{ }
\left( {\mit \Lambda}
\cS_{x_{1}^{ }\bar{\xi}_+^{ }}^{ }
- {\mit \Xi}
\cS_{x_{1}^{ }\bar{m}_+^{ }}^{ }\right) - \bar{m}_+^{ } {\mit \Xi} 
\bar{h}_{\bar{\xi}_{[+,-]}^{ }}^{ }
\cS_{\bar{m}_{[+,-]}^{ }\bar{m}_+^{ }}^{ } 
\nonumber \\& & \hspace{4.5cm} \left.
+\left({\mit \Gamma} \cS_{x_{1}^{ }\bar{m}_+^{ }}^{ }  
-\bar{m}_+^{ } {\mit \Lambda} \bar{h}_{\bar{m}_{[+,-]}^{ }}^{ }\right)
\cS_{\bar{\xi}_{[+,-]}^{ }\bar{\xi}_+^{ }}^{ } 
\right\}\,,\\
\Theta_{I}^{}&=&2\exp\{-2\beta\bar{J}_1\}-1\,,
\label{thetaI}\\
{\mit \Theta_{\rm II}^{ }} &=& 
\beta {\mit\Delta} \left[ \left(\cS_{x_{1}^{ }\bar{\xi}_+^{ }}^{ } \, 
\bar{h}_{\bar{m}_{[+,-]}^{ }}^{ } +
\cS_{x_{1}^{ }\bar{m}_+^{ }}^{ } \, 
\bar{h}_{\bar{\xi}_{[+,-]}^{ }}^{ }\right)
\cS_{\bar{m}_{[+,-]}^{ }\bar{\xi}_+^{ }}\right. \nonumber \\
&& \hspace{1.5cm} \left. -\bar{h}_{\bar{\xi}_{[+,-]}^{ }}^{ }
\cS_{x_{1}^{ }\bar{\xi}_+^{ }}^{ }
\cS_{\bar{m}_{[+,-]}^{ }\bar{m}_+^{ }}^{ } 
-\bar{h}_{\bar{m}_{[+,-]}^{ }}^{ }
\cS_{x_{1}^{ }\bar{m}_+^{ }}^{ }
\cS_{\bar{\xi}_{[+,-]}^{ }\bar{\xi}_+^{ }}^{ }
\right]\,,
\label{thetaII}
\end{eqnarray}
and
\begin{eqnarray}
\label{chide}
{\mit \Delta}&=&\left[\cS_{\bar{m}_{[+,-]}^{ }\bar{\xi}_+^{ }}^2 - 
 \cS_{\bar{\xi}_{[+,-]}^{ }\bar{\xi}_+^{ }}^{ }
 \cS_{\bar{m}_{[+,-]}^{ }\bar{m}_+^{ }}^{ }\right]^{-1}\,,\\
\label{chixi}
{\mit \Xi} &=& -S_{h\bar{\xi}_+^{ }}^{ } 
- \cS_{m_{ }^o\bar{\xi}_+^{ }}^{ } m_h^o
- \cS_{\xi_{ }^o\bar{\xi}_+^{ }}^{ } \xi_h^o\,, \\
\label{chila}
{\mit \Lambda} &=& -S_{h\bar{m}_+^{ }}^{ }
- \cS_{m_{ }^o\bar{m}_+^{ }}^{ } m_h^o
- \cS_{\xi_{ }^o\bar{m}_+^{ }}^{ } \xi_h^o\,,\\
\label{chiga}
{\mit \Gamma} &=& \bar{h}_h + \bar{h}_{m_{ }^o}^{ } m_h^o
+ \bar{h}_{\xi_{ }^o}^{ } \xi_h^o\,. 
\label{abklsg}
\end{eqnarray}
In this paper we are primiliary interested in the zero--temperature
susceptibility. Equations~(\ref{chisro})
to~(\ref{chiga}) then lead to  
\begin{equation}
\lim_{T\to 0} \chi = \left\{
\begin{array}{c@{\qquad \mbox{for} \quad}lc}
\displaystyle \chi_{\rm I}^{}  &  \bar{J}<0&\mbox{(antiferromagnetic SRO)}
\\[0.4cm]
\displaystyle \chi^o  & \bar{J}\equiv 0&\mbox{(PM)}  \\[0.4cm]
\displaystyle \frac{\chi_{\rm II}^{}}{\Theta_{\rm II}^{}}& \bar{J}> 0 
& \mbox{(ferromagnetic SRO)}.
\end{array}
\right.
\end{equation}
It is noteworthy  that, in contrast to the theory by Baumg\"artel {\it
et al.}~\cite{BSB93}, we get a finite spin susceptibility at $T=0$.
\section{Numerical results and discussion}
We have numerically solved the self--consistency equations (\ref{SROSattel1}) 
to (\ref{propA}) at $T=0$, where the SRO is included within the Bethe--Peierls
approximation, and the 2D tight--binding unperturbed density of states
corresponding to (\ref{badi}) is used. To determine the magnetic ground--state
properties (SRO versus LRO, susceptibility $\chi(0,\delta)$), all the SB
fields $b_{\alpha}$ and CVM variables ($x_{1}$, $x_{2}$) are evaluated as
functions of $U/t$ and doping. Note that  in the tedious 
numerics particular attention has to be paid to the analytical behaviour
of the complex logarithm appearing in the integrals (\ref{intSS}),
(\ref{intPaar}) and in their derivatives.

The band structure is incorporated via the non--interacting dispersion 
(\ref{badi}), where we have chosen (besides $t'=0$) the ratio $t'/t=-0.16$
according to ab--initio parameters for hole--doped ($\delta>0$) LSCO 
\cite{HSSJ90}
and ARPES data on electron--doped ($\delta<0$)
$\rm Nd_{2-\delta}Ce_{\delta}CuO_{4}$ 
\cite{Kiea93}. For both compounds, realistic values of $U/t$ lie between 
$6\lapro U/t \lapro 10$ \cite{HSSJ90}.
\subsection{Long--range order versus short--range order}
Concerning the LRO ground states of the 2D $t$--$t'$--Hubbard model, within our
{\it scalar} four--field SB approach resulting in an effective {\it Ising}
free--energy functional, only antiferromagnetic (AFM) and ferromagnetic
(FM) LRO may be described by the A--B and uniform saddle--point solutions,
respectively \cite{LMH90,DFB92}. Up to $U/t=10$ and for either doping level we
do not obtain a FM phase. This should be qualitatively contrasted with the
results obtained by Hartree--Fock approaches \cite{LH87,Br95b}, where FM states
are found for doping regions with $U/t\gapro 5$. This discrepancy may be
explained by the appearance of antiferromagnetic SRO which counteracts
ferromagnetism (as in the SRO theories of finite--temperature magnetism, cf.
Sec.~1) and which we have found to be stabilized for $U/t>5-6$ (for details
see below). The suppression of FM Hartree--Fock ground states by
antiferromagnetic SRO is in accord with Quantum Monte Carlo (QMC) 
results \cite{LH87}.

\begin{figure}[t]
\centerline{\mbox{\epsfxsize 13.0cm\epsffile{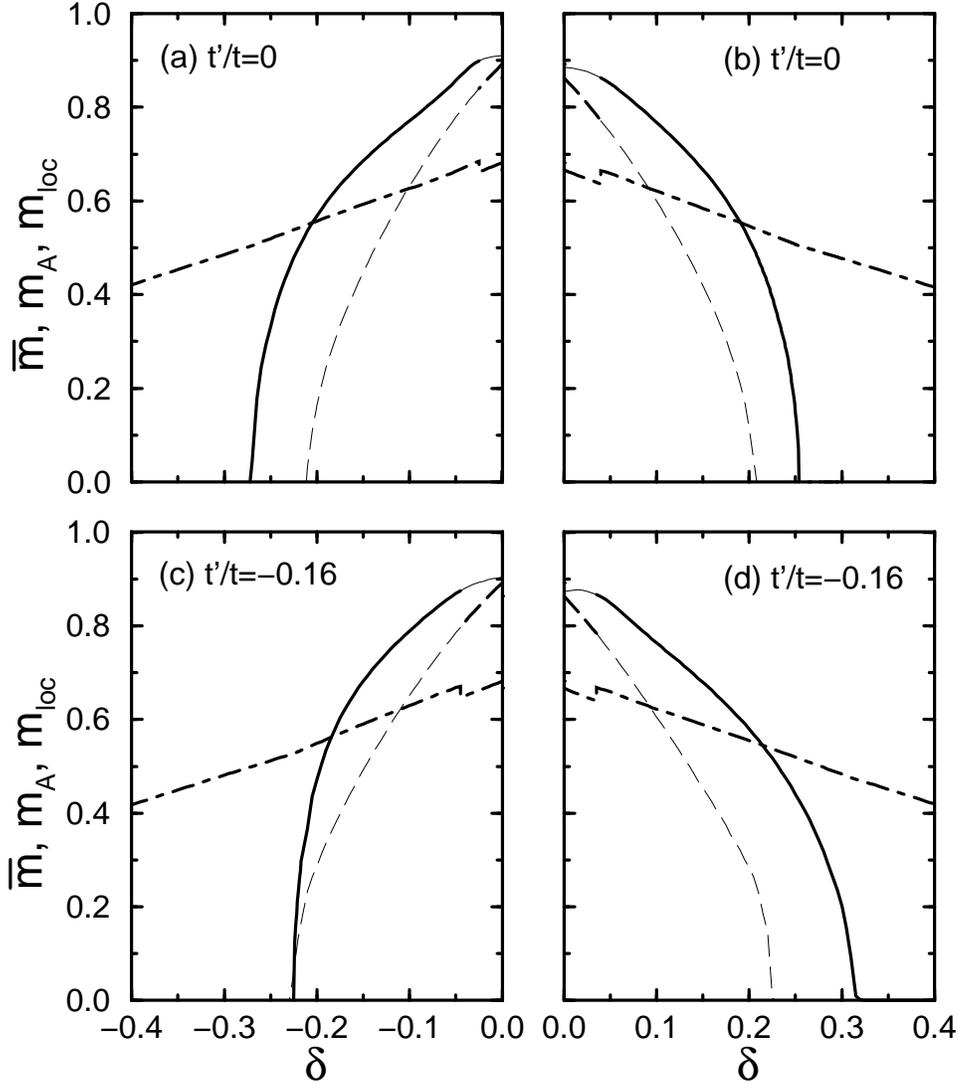}}}
\caption{Local--magnetization amplitude $\bar{m}$ (solid curve)
compared with the sublattice magnetization $m_A$ (dashed line) and the
local magnetic moment $m_{loc}$ (chain--dashed line) as functions of
electron doping ($U/t=9$; {\bf a},{\bf c}) and hole doping
($U/t=8$; {\bf b}, {\bf d}). 
Note the electron--hole symmetry for $t'=0$ and the 
asymmetry for $t'\neq 0$. The thin lines indicate the metastable solutions.}
\end{figure}
To discuss the competition between antiferromagnetic LRO and antiferromagnetic
SRO and to give more insight into our SRO concept, already described
qualitatively in Sec.~1, let us first quantify the behaviour of various
magnetic parameters as functions of hole and electron doping. 
In Fig.~2 the sublattice magnetization $m_{A}$ in the AFM phase, the
local--magnetization amplitude $\bar{m}$ being finite only in the SRO--PM
phase, where the SRO is found to  be 
antiferromagnetic ($\bar{J}_{1}<0$), and the local magnetic moment
$m_{loc}=\frac{3}{4}(n-2d^{2})$ are depicted. In the case $\delta<0$ we have
taken $U/t=9$ (instead of $U/t=8$ for $\delta>0$), in order to get an SRO--PM
phase at $t'/t=-0.16$. As can be seen, at large enough interaction
strengths we obtain an AFM$\rightleftharpoons$SRO--PM phase transition of first
order at a small critical doping $\delta_{c_{1}^{ }}$ (the thin dashed lines in
Fig.~2 indicate $m_{A}$ in the region, where the AFM solution becomes
metastable) and a
SRO--PM$\rightleftharpoons$PM transition of second order\
at $\delta_{c_{2}^{ }}$.

The local magnetic moment measures the `localization' of the electron spins
and, contrary to $\bar{m}$, is finite in the PM phase as well. With decreasing
$|\delta|$ the correlation strength increases resulting in the increase of
$m_{loc}$ (as known from the theory of itinerant magnetism). If the
`localization' prevails over the itinerancy, the formation of
antiferromagnetic SRO is energetically favoured
($|\delta|<|\delta_{c_{2}^{ }}|$). 
Once the SRO has been established, $m_{loc}$ is slightly enhanced
with respect to its value in the PM state (cf. Fig.~2).
 
\begin{figure}[t]
\centerline{\mbox{\epsfxsize 14.0cm\epsffile{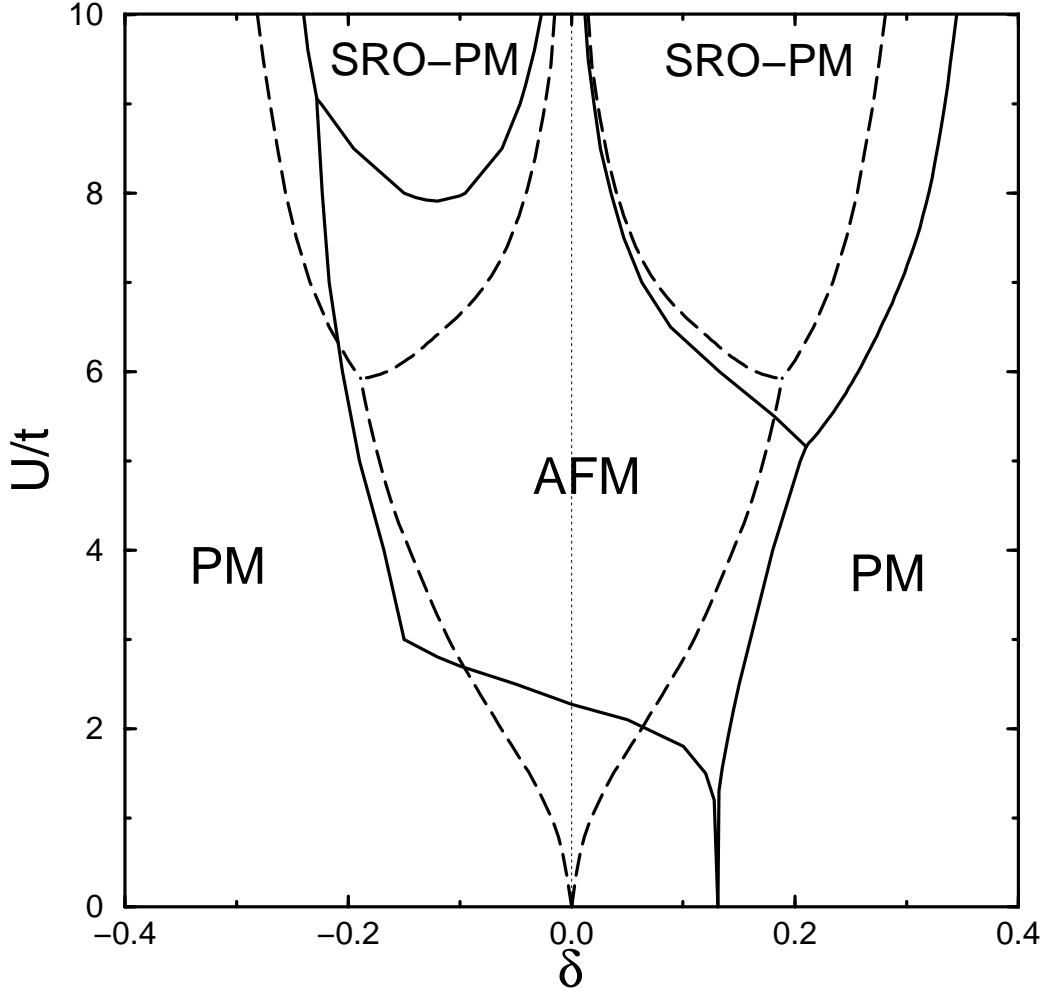}}}
\caption{Ground--state phase--diagram of the $t$--$t'$--Hubbard model,
where AFM, PM, and SRO--PM denote the antiferromagnetic LRO phase, the
paraphase without SRO, and the paraphase with SRO, respectively. The solid
(dashed) lines mark the phase boundaries at $t'/t=-0.16$ ($t'/t=0$).}
\end{figure}
In the ground--state phase diagram of the 2D $t$--$t'$--Hubbard model 
shown in Fig.~3 one observes the suppression of
LRO by SRO (at large enough $U/t$ values) in a wide doping region. The energy
comparison between the SRO--PM phase and various incommensurate spiral
LRO phases (obtained by the spin--rotation--invariant SB saddle--point
solution \cite{FW92a}) was performed in our previous work \cite{TIF95}
for the particular value  $U/t=8$, 
where also the problem of phase separation was discussed,
and yields qualitatively the same SRO effects. For $U/t=8$ (being
realistic for LSCO) and $\delta>0$ we get $\delta_{c_1^{ }}\simeq
0.04$ which agrees with the observed critical hole doping for the
destruction of antiferromagnetism in LSCO. From this result we conclude
that the rapid disappearance of antiferromagnetism in the quasi--2D
hole--doped cuprates is related to the persistence of a strong 2D
antiferromagnetic SRO in the paraphase.

For $t'/t=-0.16$ and $6\lapro U/t \lapro 8$ the pronounced
electron--hole asymmetry manifests itself in the stability of the AFM
phase in a much wider electron doping region as compared with the
hole--doped case. The $t'$--induced shrinking
of the stability region of the SRO--PM phase in the $\delta<0$ case originates
from band--structure effects and, in particular, from the shift of the
logarithmic 
Van--Hove singularity in the noninteracting density of states.
Our findings are in accord with the asymmetry of the
experimental phase diagram showing the disappearance of antiferromagnetism
in $\rm Nd_{2-\delta}Ce_{\delta}CuO_{4}$
 at $\delta_{c_1^{ }}\simeq -0.15$.

In the hole--doped systems, the hopping along the square lattice diagonals
favours antiferromagnetic correlations (Fig.~3) so that the SRO
persists up to $\delta_{c_2^{ }}\simeq 30$\% ($U/t\simeq 8$). Since
just at this doping the superconductivity in LSCO disappears, we
suggest that the antiferromagnetic SRO may be closely connected with the
superconducting pairing mechanism in the hole--doped cuprates.

Let us now discuss the weak--interaction
part of the phase diagram of the $t$--$t'$--Hubbard model, where only
the PM$\rightleftharpoons$AFM
phase transition may take place. At half--filling and $t'\neq 0$, this
transition occurs at a non--zero critical value $U_c$ rather than at
$U=0$ as for $t'=0$; at $t'/t=-0.16$ ($-0.2$) we get $U_c/t=2.3$
($2.5$). This  result agrees with the QMC value ($U_c/t=2.5\pm0.25$) 
obtained at $t'/t=-0.2$ \cite{LH87,DM96} and
may be explained by the shift  of the position of the Van--Hove singularity 
for $t'\neq 0$ (for comparison, the Hartree--Fock 
calculation yields $U_c/t=2.1$~\cite{LH87}). 
A more detailed investigation of the half--filled $t$--$t'$--Hubbard
model will be published elsewhere~\cite{TFI97}.

At the hole
doping $\delta=0.13$ and $t'/t=-0.16$ (see the phase diagram, Fig.~3), 
the AFM state is stable
against the FM state down to $U=0$
which agrees with the Hartree--Fock result of Ref.~\cite{LH87}
($\delta=0.12$ at $t'/t=-0.2$) but contradicts the result obtained by 
Brenig~\cite{Br95b}. Thus, the correct weak--interaction limit of our
saddle--point solution is confirmed. 

Let us stress again that at large interaction
strengths ($U/t>5$) the appearance of SRO, in addition to the
inclusion of an appreciable part of the correlations at the SB--PM
saddle point, changes the phase diagram qualitatively as compared with
any Hartree--Fock solution, in particular, the FM and AFM LRO states
are suppressed in a wide doping region.

\subsection{Spin susceptibility and short--range order}
Based on the solution of the self--consistency equations discussed above, in
the PM and SRO--PM phases we have evaluated the doping dependence of
the zero--temperature susceptibility $\chi(0,\delta)$ according to the
theory presented in Sec.~4.

Figure~4 shows our results for the electron--  and hole--doped cases at large
interaction strengths, where the SRO--PM phase is stabilized (cf. Fig.~3). In
the PM phase, the Pauli susceptibility of the SB--renormalized tight--binding
band reveals a pronounced increase with decreasing (electron or hole) doping
which notably depends on $t'/t$. This behaviour may be ascribed to 
Fermi--surface topology effects. Beyond the PM$\rightleftharpoons$SRO--PM
transition at $\delta_{c_{2}^{ }}$ (cf. Fig.~3), the Pauli susceptibility is
suppressed due to the antiferromagnetic--SRO--induced spin stiffness against
the orientation of the local magnetizations along the direction of 
the homogeneous external
magnetic field. Correspondingly, at $\delta_{c_{2}^{ }}$ a cusp in
$\chi(0,\delta)$ appears. According to the phase diagram, with increasing
$U/t$ the PM$\rightleftharpoons$SRO--PM transition shifts to higher doping
levels so that the peak position in $\chi(0,\delta)$ reveals the same $U/t$
dependence. Analogously, from Fig.~3 the $t'$ dependence of the peak position
emerges. With decreasing doping the degree of SRO or, equivalently, the
magnitude of the antiferromagnetic exchange energy, $|\bJone|$, increases
\cite{TIF95} so that the susceptibility decreases.

In the low--doping regime ($|\delta|\gapro |\delta_{c_{1}^{ }}|$), 
where the SRO
\begin{figure}[t]
\centerline{\mbox{\epsfxsize 14.0cm\epsffile{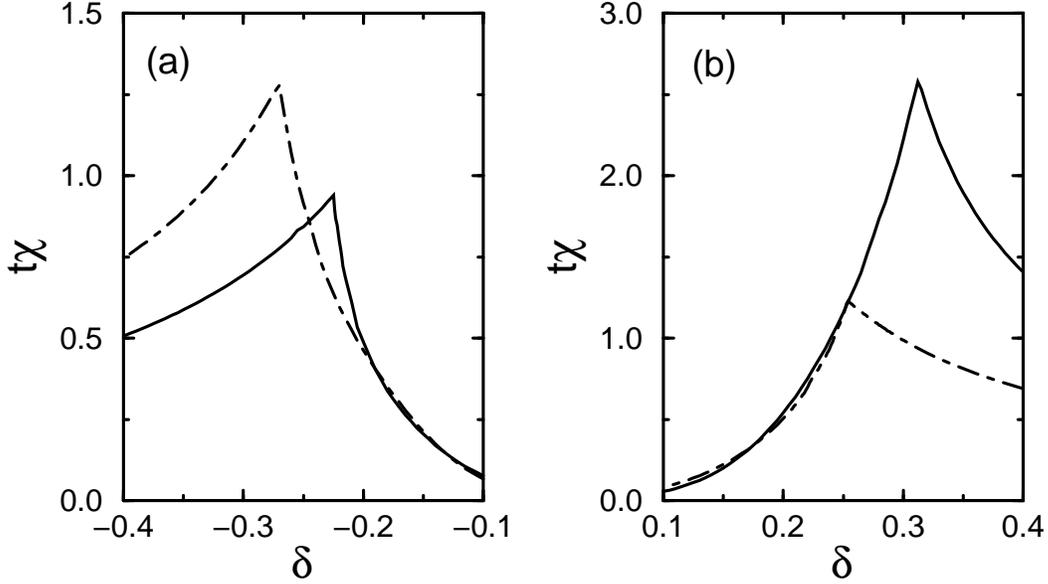}}}
\caption{Uniform static spin susceptibility at $T=0$ as a function of
electron doping ($U/t=9$; {\bf a}) and hole doping ($U/t=8$; {\bf b}).}
\end{figure}
is most pronounced and the `local' contribution to the susceptibility
(\ref{chidef}) predominates the `itinerant' contribution, the absolute value
of $\chi(0,\delta)$ may appreciably depend on the chosen approach in
describing the local properties for the following reason. In our scalar SB
approach the local properties are treated on the basis of an effective Ising
functional, describing longitudinal magnetization fluctuations only, which
yields an Ising--type `local' contribution to $\chi$. However, an improved SRO
theory based on the spin--rotation--invariant SB scheme \cite{LWH89} results
in an effective Heisenberg functional including the transverse fluctuations
too \cite{STFB96}. As known from the pure Heisenberg model \cite{Ba91}, the
finite value of the zero--temperature susceptibility (being proportional to
the inverse  exchange coupling) is due to the transverse spin
fluctuations. Therefore, we suggest that a spin--rotation--invariant theory of
SRO describing the interrelation between a Heisenberg--type `local' and
`itinerant' contribution to the susceptibility may enhance the magnitude of
$\chi$ in the low--doping region as compared with our results shown in Fig.~4.

In Fig.~5 we have depicted the zero--temperature susceptibility calculated for
the $t$--$t'$--Hubbard model taking realistic parameter sets for LSCO 
\cite{HSSJ90}
without using any fit procedure in comparison with experiments on the spin
contribution to the magnetic susceptibility of LSCO at 50 K 
\cite{Toea89,Jo89}. This
\begin{figure}[t]
\centerline{\mbox{\epsfxsize 14.0cm\epsffile{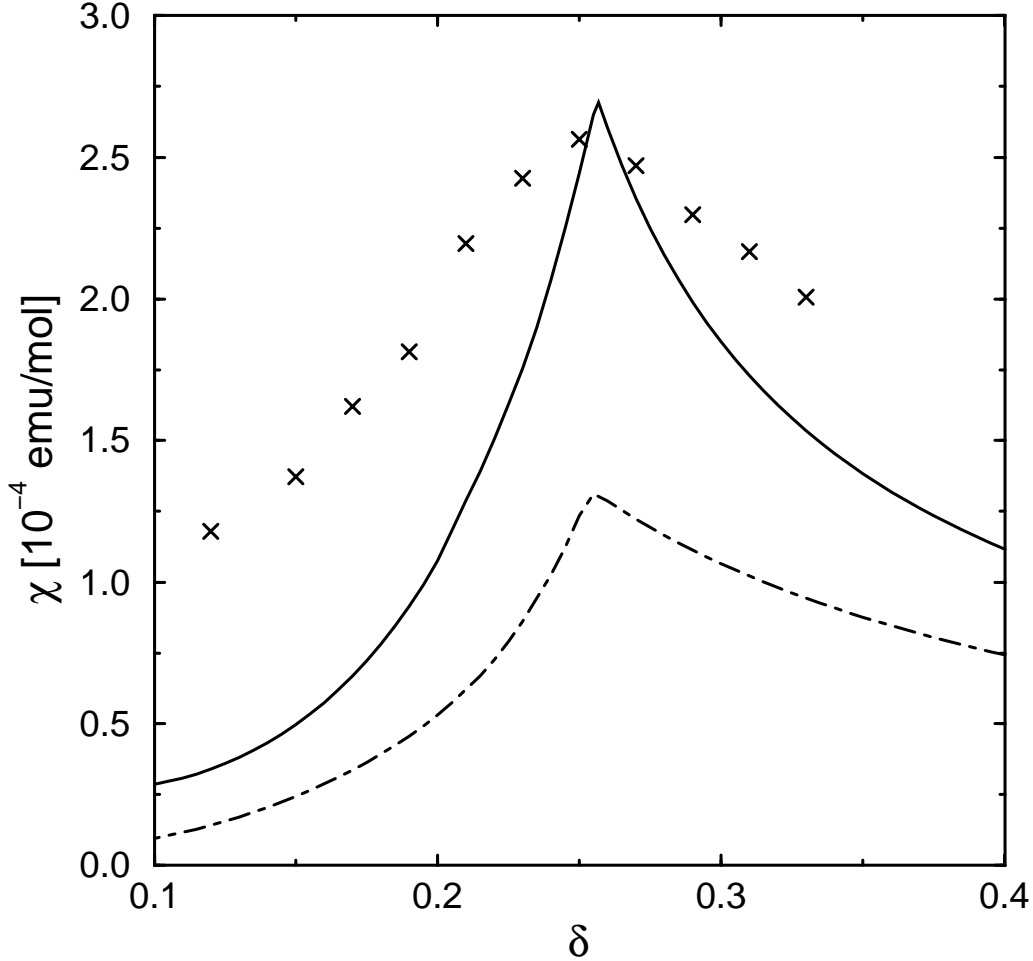}}}
\caption{Uniform static spin susceptibility at $T=0$ as a function of
hole doping. The theoretical results obtained for the 2D Hubbard model
at $U/t=6$, $t'/t=-0.16$ (solid) and $U/t=8$, $t'/t=0$
(chain--dashed), where the realistic value $t=0.3$ eV is taken, are
compared with the spin contribution ($\times$) to the (corrected)
experimental susceptibility of $\rm La_{2-\delta}Sr_\delta CuO_4$ at
$T=50$ K (Refs.~\protect\cite{Toea89} and \protect\cite{Jo89}).}
\end{figure}
contribution is obtained from the experimental data on the total
susceptibility~\cite{Toea89} by subtracting the diamagnetic core
($-9.9\times 10^{-5}$ emu/mol) and Van Vleck ($2.4\times 10^{-5}$ emu/mol)
contributions which, according to Ref. \cite{Jo89}, can be taken as
independent of doping and temperature over the limited parameter region
studied here. As Fig.~5 shows, our SRO theory reproduces very well the
position and, for suitable realistic parameters, even the height of the
observed pronounced susceptibility maximum at a hole doping of about
$25$\%. Moreover, the qualitative doping dependence of $\chi$ is in reasonable
agreement with experiments. However, in the low--doping region, the
theoretical susceptibility is too low as compared with experiments. This
deficiency may be ascribed to the neglect of transverse magnetization
fluctuations, as discussed above.

Finally, we notice that the increase of the spin susceptibility upon doping,
explained by our theory as an SRO effect in the one--band $t$--$t'$--Hubbard
model at $U/t>5$-$6$, is in qualitative accord with recent QMC data
\cite{CT94} and with results on the one--band Hubbard model ($t'=0$) based on
a semi--phenomenological approach \cite{KM94} and on the
composite--operator method \cite{MMM95}. In those works a maximum in the spin
susceptibility was found even at a smaller coupling $(U/t=4)$.

\section{Summary}
In this paper we have developed a theory of magnetic
short--range order (SRO) in itinerant electron systems on the basis of
the one--band $t$--$t'$--Hubbard model. The slave--boson
functional--integral method was improved, as compared with previous
approaches 
% \cite{MT78,Mo81,PK79,Ha81b,Ka81a,Ka91,BSB93}, 
[1, 2, 4--6, 8, 9],
by taking
into consideration the self--consistent renormalization of the
paramagnetic saddle point by SRO. The basic ingredients of our
(non--CPA) theory are the following.
\begin{itemize}
\item[(i)] The four field slave--boson functional--integral
representation is used to include an appreciable part of the
correlations already at the uniform paramagnetic (PM) saddle
point.
\item[(ii)] The SRO of local magnetizations is incorporated beyond the
PM saddle point within an effective generalized Ising model treated by
means of the cluster variational method (CVM). The CVM is illustrated
by cluster approximations with successively increasing sizes of
the basic cluster.
\item[(iii)] The SRO saddle point is determined from the free--energy
functional by minimization with respect to all Bose fields and CVM
variables.
\item[(iv)] The inclusion of a homogeneous external magnetic field at
the SRO saddle point allows the self--consistent calculation of SRO
effects on the uniform static spin susceptibility.
\end{itemize}
The theory is numerically evaluated at $T=0$, where the SRO is taken
into account at the Bethe--Peierls level of approximation. The
dependence of SRO effects on the interaction strength, the hole and
electron doping and on the band structure (modeled by $t'/t$) is
investigated. The main results are summarized as follows.
\begin{itemize}
\item[(i)] At $5<U/t<10$ ($t'/t=-0.16$) and for any doping level, the 
ferromagnetic long--range order (LRO) obtained in previous Hartree--Fock
calculations \cite{LH87} is suppressed due to the appearance of
antiferromagnetic SRO.
\item[(ii)] The antiferromagnetic and incommensurate spiral LRO phases
obtained by mean--field--type approaches make way to a paraphase with
antiferromagnetic SRO in a wide doping region. In the ground--state
phase diagram, a pronounced $t'$--induced electron--hole asymmetry is
observed.
\item[(iii)] The uniform static spin susceptibility increases upon
doping in the SRO--PM phase and shows a cusp at the transition to the
Pauli paraphase.
\end{itemize}
From the good quantitative  agreement of our theory with experiments on 
cuprates we conclude that the concept of a strong SRO in electron correlation
models for high--$T_c$ compounds may play an important role in the
explanation of many unconventional magnetic properties and of the
superconducting pairing mechanism in high $T_c$'s.
\section*{Acknowledgements}
This work was performed under the auspices of
Deutsche For\-schungsgemeinschaft under project SF--HTSL--SRO.
U.T. acknowledges the hospitality at the University of Bayreuth.
% \newpage
\begin{appendix}
\section*{Appendix A: Effective Ising model}
\reseteqn
\appAeqn
According to the reasonings in Sec.~2.2, by the use of~(\ref{Tr_I}) to
(\ref{Ansatz}) the functional (\ref{PSIO})
is expanded as
\begin{equation}
{\mit \Psi}(\{b_{\si},\si\}) = {\mit \Psi}_{ }^{(0)} 
+ {\mit \Psi}_{ }^{(1)}(\{b_{\si},\si\}) 
+ {\mit \Psi}_{ }^{(N)}(\{b_{\si},\si\})
\end{equation}
with 
\begin{equation}
{\mit \Psi}_{ }^{(0)} = \mbox{$\frac{1}{\beta}$}
\sum_{\vec{k}\sigma} \ln{\left[1-f(\Eks-\mu)\right]} 
\end{equation}
and the single--site and $N$--site fluctuation contributions,
\begin{equation}
{\mit \Psi}_{ }^{(1)}(\{b_{\si},\si\}) = \sum_{i} 
\Bigl(U\dcdsi-\nsi\nusi+\mbsi\xibsi
+\sum_{\sigma}{\mit \Phi}_{i\sigma}^{ }(b_{\si},\si) \Bigr)
\end{equation}
and
\begin{equation}
{\mit \Psi}_{ }^{(N)}(\{b_{\si},\si\}) = \int d\omega \fermi
\mbox{$\frac{1}{\pi}$}
\mbox{Im Tr}~\ln{\left[\delta_{ij}-\hGoij \Tjs(b_{\sj},\sj)\right]}\;,
\label{psin}
\end{equation}
respectively, where
\begin{equation}
\label{intSS}
{\mit \Phi}_{i\sigma}^{ }(b_{\si},\si)  = \int d\omega \fermi 
\mbox{$\frac{1}{\pi}$}
\mbox{Im}~\ln{\left[1-\hGoii \Vis(b_{\si},\si)\right]}\,.
\end{equation}
Using the identity
\begin{equation}
{\mit \Psi}(\{b_{\si},\si\}) = \sum_{\{\alpi=\pm\}}\left( \prod_i 
\frac{1+\si \alpi}{2}\right) {\mit \Psi}(\{b_{\alpi},\alpi\}) 
\end{equation}
we get 
\begin{equation}
{\mit \Psi}(\{b_{\si},\si\}) = \bPsi 
-\sum_i \bar{h}_i \si - \sum_{(ij)} 
\bar{J}_{ij} \si \sj - \sum_{(ijk)} \ldots 
\label{psikake}
\end{equation}
with 
\begin{eqnarray}
\bPsi (b_{\alpi})&=& \mbox{$\left(\frac{1}{2}\right)^N$} 
\sum\limits_{\{\alpi\}}{\mit \Psi}(\{b_{\alpi},\alpi\})\,,\\[0.1cm]
\bar{h}_i (b_{\alpi}) &=&
-\mbox{$\left(\frac{1}{2}\right)^N$} 
\sum\limits_{\{\alpi\}}\alpi
{\mit \Psi}(\{b_{\alpi},\alpi\})\,,\\[0.1cm]
\bar{J}_{ij} (b_{\alpi},b_{\alpj})&=&
-\mbox{$\left(\frac{1}{2}\right)^N$} \sum\limits_{\{\alpi\}}\alpi\alpj
{\mit \Psi}(\{b_{\alpi},\alpi\})\,.
\end{eqnarray}
Treating~(\ref{psin}) in the pair approximation we have
\begin{equation}
\mbox{Tr}~\ln{\left[\delta_{ij}
-\hGoij \Tjs(b_{\alpj},\alpj)\right]}
= \sum_{(ij)\sigma}
\ln{\left[1-\hGoij \Tjs(b_{\alpj},\alpj)\;
\hGoji \Tis(b_{\alpi},\alpi)\right]}\;,
\end{equation}
and the higher order terms in~(\ref{psikake}) vanish. Then, the
functional~(\ref{psikake}) takes the form of the generalized Ising
model~(\ref{PsiIs2}),  where the effective parameters are given by
\pagebreak
\begin{eqnarray}
\label{barPSI}
\bPsi (b_{\alpha}) &=&  {\mit \Psi}^{(0)} +\mbox{$\frac{N}{2}$}
\sum_{\alpha}\Big\{ U \dcda -\na \nua +\mba \xiba  \nonumber \\
& & \hspace*{1.5cm}
+  \sum_{\sigma}
\Big[{\mit \Phi}_{\alpha \sigma}^{ }
     +\mbox{$\frac{1}{4}$}\sum_{n} z_{n}^{ }
\left({\mit \Phi}_{n,\alpha \alpha \sigma}
     +{\mit \Phi}_{n,{-\alpha} \alpha \sigma}\right)\Big]\Big\}\;, \\
\bar{h} (b_{\alpha}) &=& -\mbox{$\frac{1}{2}$} \sum_{\alpha}\alpha
\Big[U \dcda -\na \nua +\mba \xiba
 + \sum_{\sigma} \Big( {\mit \Phi}_{\alpha \sigma}
                    +\mbox{$\frac{1}{2}$}
                   \sum_{n} z_{n}^{ }  {\mit \Phi}_{n,\alpha \alpha \sigma}
                    \Big)\Big]\;, \\
\label{JIsing}
\bar{J}_{n}^{ } (b_{\alpha}) 
&=& -\mbox{$\frac{1}{4}$}\sum_{\alpha \sigma}\left(
                 {\mit \Phi}_{n,\alpha \alpha \sigma}
                -{\mit \Phi}_{n,{-\alpha} \alpha \sigma} \right) \;.
\end{eqnarray}
Here, ${\mit \Phi}_{\alpha \sigma}^{ }=\left.{\mit \Phi}_{i\sigma}^{
}(\alpi)\right|_{\alpi=\alpha}$, and  
${\mit \Phi}_{n,\alpha \alpha^{\prime}\sigma}=\left.
{\mit \Phi}_{(ij)_{n}^{} \sigma}(\alpi,\alpj)
\right|_{{{\alpi=\alpha}^{ }\atop {\alpj=\alpha^{\prime}}}}$
with
\begin{equation}
\label{intPaar}
{\mit \Phi}_{(ij)_{n}^{}\sigma}^{ } = \int d\omega \fermi 
\mbox{$\frac{1}{\pi}$}
\mbox{Im}~\ln\left[1-\hat{G}^o_{(ij)_{n}^{} \sigma}
\Tjs(b_{\alpj},\alpj)\,\hat{G}^o_{(ji)_{n}^{} \sigma} 
\Tis(b_{\alpi},\alpi)\right]
\end{equation}
is the two--site fluctuation contribution which takes into account 
SRO effects.

\section*{Appendix B: Cluster variational method and examples}
\reseteqn
\appBeqn
To illustrate the algorithm of the CVM sketched in Sec.~2.3, we
consider the simplest cluster approximations with the basic clusters
$(a)$, $(b)$ and $(c)$ of Fig.~1 corresponding to the
Bragg--Williams~\cite{BW34}, Bethe--Peierls~\cite{Be35} and 
Kramers--Wannier~\cite{KW41} approximations, respectively. The energies
$\varepsilon_\kappa^{(v)}$ and their multiplicities
$\lambda_\kappa^{(v)}$ for the topologically inequivalent spin
configurations of the isolated clusters are given in Table~1. The
\begin{table}[ht]
 \centerline{
\begin{tabular}{|c||c|c|r||c|c|r||c|c|r|}
\cline{2-10}
%---------------------------------------------------------------
\multicolumn{1}{c}{\rule[-3mm]{0mm}{8mm}} &
\multicolumn{3}{|c||}{\boldmath$(a)$ =\raisebox{-0.2cm}{\usebox{\heins}}} &
\multicolumn{3}{|c||}{\boldmath$(b)$ =\raisebox{-0.2cm}{\usebox{\hzwei}}} &
\multicolumn{3}{|c|}{\boldmath$(c)$ =\raisebox{-0.2cm}{\usebox{\hvier}}} \\ 
\hline
%---------------------------------------------------------------
$\kappa$ &
\rule[-0.2cm]{0cm}{0.8cm}  &
$\lambda_{\kappa}^{(a)}$ &
$\varepsilon_{\kappa}^{(a)}$ &
\rule[-0.2cm]{0cm}{0.8cm} &
$\lambda_{\kappa}^{(b)}$ &
$\varepsilon_{\kappa}^{(b)}$\hspace{0.6cm}  &
\rule[-0.2cm]{0cm}{0.8cm} &
$\lambda_{\kappa}^{(c)}$ &
$\varepsilon_{\kappa}^{(c)}$\hspace{1.3cm} \\ 
\hline \hline
%---------------------------------------------------------------
1 &
\raisebox{-0.3cm}{\mbox{\usebox{\einsp}}} &
1 &
$-\bh$ &
\raisebox{-0.3cm}{\mbox{\usebox{\zweipp}}} &
1 &
$-2\bar{h}-\bar{J}_{1}$ &
\raisebox{-0.3cm}{\mbox{\usebox{\vierpppp}}} &
1 &
$-4\bh-4\bar{J}_{1}-2\bar{J}_{2}$ \\ \hline
%---------------------------------------------------------------
2 &
\raisebox{-0.3cm}{\mbox{\usebox{\einsm}}} &
1 &
$\;\;\,\bar{h}$ &
\raisebox{-0.3cm}{\mbox{\usebox{\zweipm}}} &
2 &
$\bar{J_{1}}$ &
\raisebox{-0.3cm}{\mbox{\usebox{\vierpppm}}} &
4 &
$-2\bh$\hspace*{2.1cm}\\ \hline
%---------------------------------------------------------------
3 &
\multicolumn{3}{|c||}{}&
\raisebox{-0.3cm}{\mbox{\usebox{\zweimm}}} &
1 &
$+2\bh-\bar{J}_{1}$ &
\raisebox{-0.3cm}{\mbox{\usebox{\vierppmm}}} &
4 & 
$2\bar{J}_{2}$ \\ \cline{1-1}\cline{5-10}
%---------------------------------------------------------------
4 &
\multicolumn{3}{|c||}{}&
\multicolumn{3}{|c||}{}&
\raisebox{-0.3cm}{\mbox{\usebox{\vierpmpm}}} &
2 & 
$4\bar{J}_{1}-2\bar{J}_{2}$ \\ \cline{1-1}\cline{8-10}
%---------------------------------------------------------------
5 &
\multicolumn{3}{|c||}{}&
\multicolumn{3}{|c||}{}&
\raisebox{-0.3cm}{\mbox{\usebox{\vierpmmm}}} &
4 & 
$2\bh$\hspace*{2.1cm} \\ \cline{1-1}\cline{8-10}
%---------------------------------------------------------------
6 &
\multicolumn{3}{|c||}{}&
\multicolumn{3}{|c||}{}&
\raisebox{-0.3cm}{\mbox{\usebox{\viermmmm}}} &
1 & 
$4\bh-4\bar{J}_{1}-2\bar{J}_{2}$ \\ \hline
\end{tabular}}

 \caption{The topologically different configurations
 $\kappa$ ($\bullet=\uparrow$, $\circ=\downarrow$), multiplicities 
 $\lambda_{\kappa}^{(v)}$ and configuration energies 
 $\varepsilon_{\kappa}^{(v)}$ for the clusters $(a)$, $(b)$ and $(c)$.}
\end{table}
consistency relations~(\ref{konrel}) between the distribution numbers
$p_\kappa^{(v)}$ of the subclusters and the basic clusters 
$p_\kappa^{(w)}$ as well as the dependences of $p_\kappa^{(w)}$ on the
mutually independent CVM variables $x_\iota^{ }$ for the $(w)$--clusters
are listed in Table~2. Concerning the choice of the variables
$x_\iota^{ }$ in proceeding from the $(w)$--cluster to the $(w+1)$--cluster
approximation (e.g. $(a)\to(b)$), it is convenient to preserve the
definitions of the $x_{\iota}^{ }$ or, equivalently, the $x$--dependences
of $p_\kappa^{(v)}$ with $(v)\subseteq(w)$. Then, only the $x$--dependences of
$p_\kappa^{(w+1)}$ have to be determined from the consistency
relations.
Accordingly, we take the choice  $x_{1}^{ }=p_{1}^{(a)}-p_{2}^{(a)}$,  
$x_{2}^{ }=p_{2}^{(b)}$, $x_{3}^{ }=p_{3}^{(c)}$ , 
$x_{4}^{ }=p_{4}^{(c)}$ and $x_{5}^{ }=p_{2}^{(c)}-p_{5}^{(c)}$.

As can be seen from Table~2, the expectation value $\Kom=\langle \si
\rangle$ is given by $\Kom=x_1^{ }$ (being the order parameter in the
ferromagnetic Ising model).

\begin{table}[ht]
 \centerline{
\begin{tabular}{|l||l|l|}
\hline
\rule[-3mm]{0mm}{8mm} Cluster &
consistency relations &
CVM variables $x_{\iota}$\\ \hline\hline
$(c)$ &
\rule[-3mm]{0mm}{8mm}
$p_{1}^{(c)}$ &
$p_{1}^{(c)}=\frac{1}{2}(1+x_{1}^{ }-4x_{2}^{ }+2x_{3}^{ }-2x_{5}^{ })$ 
\\ \cline{2-3}
&
\rule[-3mm]{0mm}{8mm}
$p_{2}^{(c)}$ &
$p_{2}^{(c)}=\frac{1}{2}(x_{2}^{ }-x_{3}^{ }-x_{4}^{ }+x_{5}^{ })$ 
\\ \cline{2-3}
&
\rule[-3mm]{0mm}{8mm}
$p_{3}^{(c)}$ &
$p_{3}^{(c)}=x_{3}^{ }$ 
\\ \cline{2-3}
&
\rule[-3mm]{0mm}{8mm}
$p_{4}^{(c)}$ &
$p_{4}^{(c)}=x_{4}^{ }$ 
\\ \cline{2-3}
&
\rule[-3mm]{0mm}{8mm}
$p_{5}^{(c)}$ &
$p_{5}^{(c)}=\frac{1}{2}(x_{2}^{ }-x_{3}^{ }-x_{4}^{ }-x_{5}^{ })$ 
\\ \cline{2-3}
&
\rule[-3mm]{0mm}{8mm}
$p_{6}^{(c)}$ &
$p_{6}^{(c)}=\frac{1}{2}(1-x_{1}^{ }-4x_{2}^{ }+2x_{3}^{ }+2x_{5}^{ })$ 
\\ \hline\hline
$(b)$ &
\rule[-3mm]{0mm}{8mm}
$p_{1}^{(b)}=p_{1}^{(c)}+2p_{2}^{(c)}+p_{3}^{(c)}$ &
$p_{1}^{(b)}=\frac{1}{2}(1+x_{1}^{ }-2x_{2}^{ })$ 
\\ \cline{2-3}
&
\rule[-3mm]{0mm}{8mm}
$p_{2}^{(b)}=p_{2}^{(c)}+p_{3}^{(c)}+p_{4}^{(c)}+p_{5}^{(c)}$ &
$p_{2}^{(b)}=x_{2}^{ }$ 
\\ \cline{2-3}
&
\rule[-3mm]{0mm}{8mm}
$p_{3}^{(b)}=p_{3}^{(c)}+2p_{5}^{(c)}+p_{6}^{(c)}$ &
$p_{3}^{(b)}=\frac{1}{2}(1-x_{1}^{ }-2x_{2}^{ })$ 
\\ \hline\hline
$(a)$ &
\rule[-3mm]{0mm}{8mm}
$p_{1}^{(a)}=p_{1}^{(b)}+p_{2}^{(b)}$ &
$p_{1}^{(a)}=\frac{1}{2}(1+x_{1}^{ })$ 
\\ \cline{2-3}
&
\rule[-3mm]{0mm}{8mm}
$p_{2}^{(a)}=p_{2}^{(b)}+p_{3}^{(b)}$ &
$p_{2}^{(a)}=\frac{1}{2}(1-x_{1}^{ })$ 
\\ \hline
\end{tabular}}

 \caption{Distribution numbers $p_{\kappa}^{(v)}$ 
  of the configurations $\kappa$ (Table~1) as functions of
  the distribution numbers of the basic cluster $(w)$
  (consistency relations) and as functions of the CVM variables
  $x_\iota$. Note the hierarchy in the choice of $x_\iota$ (see text).}
\end{table}
The geometrical factors $y_w^{(v)}$appearing in eq.~(\ref{fhb1}) are
defined by the set of equations~\cite{HB55}
\begin{equation}
\label{gfhb}
\sum_{v=a}^w \alpha_{ }^{(u)(v)}y_w^{(v)}=x_{ }^{(u)}\;;
\;\;\;(u)\subseteq(w)\;,
\end{equation}
where the matrix element $\alpha_{ }^{(u)(v)}$ indicates the number of
clusters $(u)$ contained in the cluster $(v)$ and $x_{ }^{(u)}$ gives
the total number of $(u)$--clusters in the lattice divided by $N$.

Let us now exemplify the CVM in the course of successively increasing
sizes of the basic cluster $(w)$.
\begin{itemize}
\item[(i)] Bragg--Williams approximation\\
Taking a single site as basic `cluster', i.e. $(w)=(a)$, from
Tables~1 and 2 we obtain the free energy~(\ref{fhb2}) as
\begin{equation}
\label{f2bw}
\cF_{ }^{(a)}(x_{1})=-N\left[x_{1}\bh
-\frac{1}{\beta}\sum_{\kappa=1}^{2}p_{\kappa}^{(a)}\ln
p_{\kappa}^{(a)} \right]\;.
\end{equation}
Since there are no subclusters and the total number of $(a)$--clusters
in the lattice coincides with $N$ ($x_{ }^{(a)}=1$), from (\ref{gfhb})
we get trivially $y_a^{(a)}=1$. Thus, the free energy (\ref{fhb1}) in
the Bragg--Williams approximation is
\begin{equation}
\cF_{a}^{ }(x_{1})=\cF_{ }^{(a)}(x_{1})\;.
\end{equation}
The CVM equations (\ref{CVMGl}) reduce to the equation 
$\frac{1}{N}\frac{\partial \cF_a}{\partial x_1}=0$, which, in the
paraphase and the $h=0$ limit, has the solution $x_{1}^{ }=0$.

\item[(ii)] Bethe--Peierls approximation\\
Choosing the cluster $(b)$ as basic cluster (Table~1), the number of
relevant spin configurations and CVM variables (Table~2) increases by
one. Then, the free energy (\ref{fhb2}) of an assembly of $N$
independent Bethe clusters $(b)$ is obtained as
\begin{equation}
\label{f2bp}
\cF_{ }^{(b)}(x_{1},x_{1})=-N\left[x_{1}^{ }\bh+(1-4 x_{2})\bar{J}_{1}^{ }
-\frac{1}{\beta}\sum_{\kappa=1}^{3}p_{\kappa}^{(b)}\ln 
p_{\kappa}^{(b)}\right]\;.
\end{equation}
Determining the geometrical factors $y_w^{(v)}$ from (\ref{gfhb}), we
note that the cluster $(a)$ is contained twice in $(b)$, and the
cluster $(b)$ is contained in the lattice $2N$ times. Accordingly, the
equations (\ref{gfhb}) become
\begin{equation}
\label{gfbp}
 \left(\!\!\! \begin{array}{cc}
           1 &  2 \\
           0 &  1 
        \end{array}\!\!\!\right)
 \left(\!\!\!\begin{array}{c}
            y_{b}^{(a)}\\
            y_{b}^{(b)}
        \end{array}\!\!\!\right)=
 \left(\!\!\! \begin{array}{c}
            1\\
            2
        \end{array}\!\!\!\right)\;.
\end{equation}
From the solution of (\ref{gfbp}) and by (\ref{fhb1}) we obtain the
free energy in the Bethe--Peierls approximation,
\begin{equation}
\cF_{b}^{ }(x_{1},x_{2})=2 \cF_{ }^{(b)}-3 \cF_{ }^{(a)}\;,
\end{equation}
with $\cF_{ }^{(b)}$ and $\cF_{ }^{(a)}$ given by (\ref{f2bp}) and
(\ref{f2bw}), respectively.

By (\ref{korrf}), the nearest--neighbour pair correlation function
$\Konem=\langle \si\sj \rangle_1^{ }$ is expressed as
\begin{equation}
\label{NNkorr}
\Konem=1-4 x_{2}\;.
\end{equation}

Concerning the CVM equations 
$\frac{1}{N}\frac{\partial \cF_b}{\partial x_\iota}=0$ ($\iota=1,2$),
the equation for $x_1$ reads
\begin{equation}
\label{BPcvm1}
-\bh
+\frac{1}{\beta}\left[\frac{3}{2}\ln{\frac{1-x_{1}}{1+x_{1}}}
	              +\ln{\frac{1+x_{1}-2x_{2}}{1-x_{1}-2x_{2}}}\right]  
= 0\;.
\end{equation}
In the paraphase, eq.~(\ref{BPcvm1}) has a non--trivial solution for
$h\neq0$ only and therefore is needed in the calculation of the spin
susceptibility (Sec.~4). In the $h=0$ limit ($x_{1}=0$) there remains
the second CVM equation\begin{equation}
2 x_{2} I_{1}^{2}=1-2 x_{2}
\end{equation}
with
\begin{equation}
\label{Idef}
I_{n}:=\exp{\{\beta\bar{J}_{n}\}}\;.
\end{equation}
The solution gives
\begin{equation}
\label{BPsol}
x_{2}=\mbox{$\frac{1}{4}$}[1-\tanh{(\beta \bar{J}_{1})}]\;.
\end{equation}
%--------------------------------------------------------------------------
Inserting (\ref{BPsol}) into (\ref{NNkorr}) one gets the same correlation
function $\Konem$ as in the original Bethe cluster method (working
with an effective Bethe field).

\item[(iii)] Kramers--Wannier approximation\\
Here, the cluster $(c)$ is taken as a basic cluster, and the number of
CVM variables is enlarged by three (Tables~1,~2). The free energy
(\ref{fhb2}) for $(v)=(c)$ is given by 
\begin{eqnarray}
\label{FKW}
\cF^{(c)}(x_{1},\ldots,x_{5}) &=& -N\Bigl\{4x_{1}\bh+4(1-4 x_{2})\bar{J}_{1}
\nonumber \Bigr.\\
& & \hspace*{-0.5cm}\Bigl.+2[1-4(x_{2}+x_{3}-x_{4})]\bar{J}_{2}
-\frac{1}{\beta}\sum_{\kappa=1}^{6}\lambda_{\kappa}^{(c)} p_{\kappa}^{(c)}
\ln p_{\kappa}^{(c)}\Bigr\}\;.
\end{eqnarray}

From (\ref{gfhb}) we obtain the geometrical factors $y_w^{(v)}$ as
solutions of the equations  
\begin{equation}
\label{gfkw}
 \left(\!\!\! \begin{array}{ccc}
           1 & 2 & 4\\
           0 & 1 & 4\\
           0 & 0 & 1 
        \end{array}\!\!\!\right)
 \left(\!\!\!\begin{array}{c}
            y_{c}^{(a)}\\
            y_{c}^{(b)}\\
            y_{c}^{(c)}
        \end{array}\!\!\!\right)=
 \left(\!\!\! \begin{array}{c}
            1\\
            2\\
            1
        \end{array}\!\!\!\right)\;,
\end{equation}
which yield the free energy in the Kramers--Wannier approximation, 
\begin{equation}
\label{fKW}
\cF_{c}^{ }(x_{1},\ldots,x_{5})= \cF_{ }^{(c)}
-2 \cF_{ }^{(b)}+\cF_{ }^{(a)}\;.
\end{equation}
Besides the nearest--neighbour correlation function (\ref{NNkorr}), 
from (\ref{korrf}), (\ref{FKW}) and (\ref{fKW}) the next--nearest--neighbour
correlation function 
\begin{equation}
K_{\bar{J}_2}^{ }=1-4(x_{2}+x_{3}-x_{4})
\end{equation}
can be calculated. In the paraphase and in the $h=0$ limit, the CVM
equations yield $x_{1}=x_{5}=0$, and the remaining equations become
\begin{equation}
\label{KWcvm}
p_{4}^{(c)}p_{1}^{(b)}I_{1}^{2}=p_{2}^{(c)}p_{2}^{(b)} \;,\;\;\;
p_{3}^{(c)}I_{2}^{2}=p_{2}^{(c)} \;,\;\;\;
p_{1}^{(c)}p_{3}^{(c)}p_{4}^{(c)}=p_{2}^{(c)2} \;
\end{equation}
with the distribution numbers taken from Table~2 and the definitions
(\ref{Idef}). Instead of giving the explicit solutions of (\ref{KWcvm}),
we sketch the resulting zero--temperature behaviour in
Fig.~6. There, the degenerate ground--state spin
configurations for $h=0$ are shown in the $\bar{J}_1$--$\bar{J}_2$
plane. For $\bar{J}_2=0$, we have antiferromagnetic $(\pi,\pi)$  SRO if
$\bar{J}_1<0$ and ferromagnetic $(0,0)$ SRO if $\bar{J}_1>0$. The
inclusion of the next--nearest--neighbour Ising exchange energy
$\bar{J}_2$ gives rise to frustration effects and permits to describe
SRO of a stripe structure denoted by $(\pi,0)$--SRO.\\  
\begin{figure}[ht]
% \begin{minipage}[b]{8cm}
\centerline{
\unitlength1.0mm
\begin{picture}(70,55)
{\thicklines
\put(35,25){\line( 2,-1){35}}
\put(35,25){\line(-2,-1){35}}
\put(35,25){\line( 0, 1){20}}}
\put( 0,25){\vector(1,0){70}}
\put(35, 0){\vector(0,1){50}}
\put(66,27){$\bar{J}_{1}$}
\put(30,47){$\bar{J}_{2}$}
\put( 2,17){$\bar{J}_{2}=\bar{J}_{1}/2$}
\put(55,17){$\bar{J}_{2}=-\bar{J}_{1}/2$}
\put(10,40){$(\pi,\pi)-{\rm SRO}$}
\put(10,28){\usebox{\vierpmpm}}
\put(20,28){\usebox{\viermpmp}}
\put(40,40){$(0,0)-{\rm SRO}$}
\put(40,28){\usebox{\vierpppp}}
\put(50,28){\usebox{\viermmmm}}
\put(26,12){$(\pi,0)-{\rm SRO}$}
\put(13, 1){\usebox{\vierppmm}}
\put(24, 1){\usebox{\vierpmmp}}
\put(37, 1){\usebox{\viermmpp}}
\put(48, 1){\usebox{\viermppm}}
\end{picture}}
%--------------------------------------------------------------------------
\label{KWLoes}
\caption{Paramagnetic solutions of the CVM equations at
$T=0$ within the Kramers--Wannier approximation.}
% \end{minipage}
\end{figure}
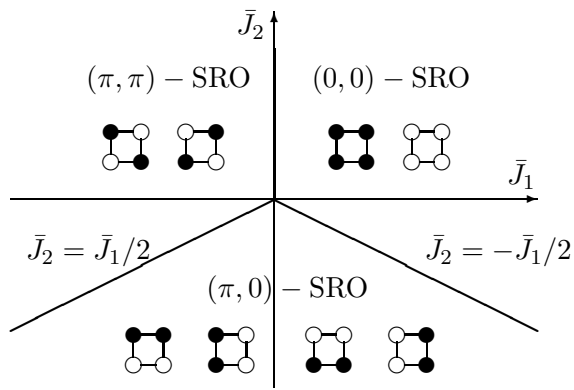
\end{itemize}

%\section*{Appendix C: Susceptibility -- matrix elements}
%\reseteqn
%\appCeqn
%\input{ta3}
\end{appendix}
\newpage
\baselineskip0.62cm
\bibliography{ref}
\bibliographystyle{phys}
\newpage
%\figure{Fig.~1: Uniform static spin susceptibility}
%\centerline{\mbox{\epsfxsize 15cm\epsffile{susz-T0.eps}}}
\end{document}